\documentclass[12pt]{iopart}
\usepackage[utf8]{inputenc}
\usepackage{graphicx}
\usepackage{braket}
\usepackage{mathtools}
\usepackage[colorlinks=true,citecolor=blue,linkcolor=blue,urlcolor=blue]{hyperref}
\usepackage{cite}

\graphicspath{{figures/}} 

\newcommand{\average}[1]{\left\langle#1\right\rangle}

\newcommand\redsout{\bgroup\markoverwith{\textcolor{red}{\rule[0.5ex]{2pt}{0.4pt}}}\ULon}
\newcommand\rhodm{\hat{\rho}}

\newcommand{\Ek}{\hat{E}_k}
\newcommand{\Ekc}{\hat{E}_k^\dagger}
\newcommand{\ketS}[1]{\ket{#1}_{\!\!{}_\mathcal{S}}}

\newcommand{\ketR}[1]{\ket{#1}_{\!\!{}_\mathcal{R}}}
\newcommand{\braS}[1]{{}_{{}_\mathcal{S}}\!\!\bra{#1}}

\newcommand{\braR}[1]{{}_{{}_\mathcal{R}}\!\!\bra{#1}}
\newcommand{\ketbraS}[2]{\ket{#1}_{\!{}_\mathcal{S}}\!\!\bra{#2}}
\newcommand{\ketbraR}[2]{\ket{#1}_{\!{}_\mathcal{R}}\!\!\bra{#2}}

\newcommand{\dS}{d_{{}_\mathcal{S}}}

\newcommand{\dR}{d_{{}_\mathcal{R}}}

\usepackage[super]{nth}

\begin{document}

\title{Quantum transfer of interacting qubits}

\author{Tony J. G. Apollaro}
\ead{tony.apollaro@um.edu.mt}
\address{Department of Physics, University of Malta, Msida MSD 2080, Malta.}

\author{Salvatore Lorenzo}
\address{Universit\`a degli Studi di Palermo, Dipartimento di Fisica e Chimica - Emilio Segr\`e, via Archirafi 36, I-90123 Palermo, Italy}

\author{Francesco Plastina}
\address{Dipartimento di Fisica, Universit\`a della Calabria, 87036 Arcavacata di Rende (CS), Italy.}
\address{INFN, gruppo collegato di Cosenza.}

\author{Mirko Consiglio}
\address{Department of Physics, University of Malta, Msida MSD 2080, Malta.}

\author{Karol \.{Z}yczkowski}
\address{Institute of Theoretical Physics, 
Jagiellonian University, ul. {\L}ojasiewicza 11, 30--348 Krak\'ow, Poland.}
\address{Center for Theoretical Physics, Polish Academy of Sciences, Al. Lotnik\'{o}w 32/46, 02-668 Warszawa, Poland.}

\date{\today{}}

\begin{abstract}
The transfer of quantum information between different locations is key to many quantum information processing tasks. Whereas, the transfer of a single qubit state has been extensively investigated, the transfer of a many-body system configuration has insofar remained  elusive. We address the problem of transferring the state of $n$ \textit{interacting} qubits. Both the exponentially increasing Hilbert space dimension, and the presence of interactions significantly scale-up the complexity of achieving high-fidelity transfer. By employing tools from random matrix theory and using the formalism of quantum dynamical maps, we derive a general expression for the average and the variance of the fidelity of an arbitrary quantum state transfer protocol for $n$ interacting qubits. Finally, by adopting a weak-coupling scheme in a spin chain, we obtain the explicit conditions for high-fidelity transfer of 3 and 4 interacting qubits.
\end{abstract}

\maketitle

\section{Introduction}
Quantum Information Processing (QIP) is shaping the \nth{21} century technology by means of the advantage it provides, with respect to its classical counterpart, in fields ranging from computation to cryptography~\cite{Nielsen2010}. A basic building block of several QIP protocols is the transfer of quantum information between different locations. In particular, a great variety of different protocols have been devised to achieve the high fidelity transfer of the quantum state of a single qubit (1-QST). They can be classified into three broad classes: protocols employing flying qubits~\cite{Northup2014}, teleportation-based ones~\cite{Bennett1993}, and those employing spin-$\frac{1}{2}$ chains as quantum data bus~\cite{Bose2003}.
In this paper, we will focus on the last approach. Since the seminal paper of Bose \cite{Bose2003}, numerous 1-QST protocols have been proposed, and implemented, in systems ranging from evanescently coupled optical waveguides~\cite{Tamascelli2016, Chapman2016a} and cavity-coupled atoms and ions~\cite{Vogell2017a}, to transmon qubits~\cite{Li2018a}, nitrogen vacancy centers~\cite{Ali2018} and Rydberg atoms~\cite{Dlaska2017}, just to name a few.

As a direct generalization of the original set-up,  the quantum state transfer of $n$-qubits ($n$-QST) has received some attention in the last few years, with a focus on the transfer of entangled qubit states. In the majority of cases, however, the state to be sent is that of a set of non-interacting qubits~\cite{Vieira2019, Vieira2020, Yousefjani2020, Yousefjani2020a}. On the other hand, the transfer of the quantum state of \textit{interacting} $n$-body systems has been insofar barely addressed.
In fact, for many QIP protocols, fast and efficient non-interacting $n$-QST would already constitute an important achievement: in distributed quantum computing architectures~\cite{Preskill2018}, for instance, the $n$-qubit output state of a computation has to be distributed among different quantum processors. In quantum secret sharing protocols an entangled $n$-qubit state is shared among several users~\cite{Ouyang2017a}, while a fully-fledged quantum internet~\cite{Wehner2018} requires nodes capable of exchanging, ideally, arbitrary many-body states.
However, the quantum state transfer of interacting qubits ($n$-iQST) could be far more beneficial both for QIP protocols (where, e.g., qubit-decoupling operations prior to transfer could be avoided), and, in perspective, to also achieve the transfer of the full physical configuration of complex systems, where complexity is embodied not only by the dimensionality of the system's Hilbert space~\cite{Society2004} but also by the interactions among its constituents~\cite{Anderson1972}. The transfer of an interacting system's state may also constitute a significant advantage in those experiments with many-body systems, where state preparation is not easily feasible in the same set-up in which the rest of the experiment is performed. In these cases, one could, e.g., prepare the state embodying the properties under investigation in a different location and send it later, via a quantum channel, in order to load it into the main set-up.

Although being closely related, $n$-QST and $n$-iQST represent two different aspects of quantum transfer, with the latter being a non-trivial generalisation of the former.
In the non-interacting case, indeed, the $n$-qubit state could in principle be transferred sequentially, with qubits sent one by one. On the other hand, if we aim at transferring the quantum state while the system is interacting, sequential transfer is not feasible as it has to be accomplished simultaneously for the entire system. Similarly to $n$-QST, for $n$-iQST, the recipient is required to have a physical system able to store the received quantum information. In our case, it is sufficient to work with a qubit system with the same interaction scheme of the one that is sent. Therefore, the quality of a $n$-iQST can be assessed with the same figure of merits as those used for $n$-QST; namely, the fidelity.

The $n$-QST problem has been posed shortly after 1-QST via spin-$\frac{1}{2}$ chain was proposed. Following the protocol for perfect state transfer (PST), which entails fully-engineered couplings of the quantum chain, it has been shown that mirror-periodic Hamiltonians allow for mirror-inversion of an arbitrary quantum state with respect to the center of the chain~\cite{Albanese2004, Kay2010}. However, in order to apply the same idea to achieve $n$-iQST, one would require a fully-engineered quantum channel, implying, in general, a modification of the couplings among the qubits embodying the sender's system. On the other hand, uniformly-coupled chains allow for high-quality $n$-QST for specific lengths related to prime number theory~\cite{Sousa2014a}, and extensive research has been devoted to investigate the transfer of few-qubit entangled states over such spin chains~\cite{Lorenz2014,Yousefjani2020,Yousefjani2020a,Apollaro2020,Vieira2020, VIEIRA20182586,Almeida2019}.
However, a general approach to the $n$-QST problem has not been put forward yet, and, moreover, specific $n$-QST protocols may not be applicable to the $n$-iQST case due to the dynamical evolution of the sender's qubits.


Because of the exponential increase of the Hilbert space dimension $d=2^n$ of an $n$-qubit state, QST protocols based on LOCC (local operations and classical communication) perform poorly, displaying a maximal fidelity scaling as $\frac{2}{d+1}$~\cite{Horodecki1999}. It is, therefore, of the utmost importance to identify the conditions on a quantum dynamical map that allows it to surpass the LOCC limit of $n$-QST, and to single out physical models that realise these maps.

In this paper, we address the $n$-iQST problem via a novel approach, by combining quantum dynamical maps~\cite{Society2004} and random matrix theory~\cite{Aubert2004}, in order to derive the average fidelity, and its variance, of an arbitrary $n$-iQST protocol. Our approach will be similar to that employed in Ref.~\cite{Bayat2007} to determine the average fidelity for the QST of a qudit. We find that the $n$-iQST average fidelity can be decomposed into two contributions. A first, classical one achieving the LOCC limit, resulting only from map elements connecting the diagonal elements of the sender and receiver density matrices, and a second, quantum contribution embodied by the map elements connecting off-diagonal density matrix elements of the sender and the receiver. Finally, we apply our information-theoretical formalism to the $n$-iQST via a spin-$\frac{1}{2}$ chain obtaining the conditions for high-quality iQST via a uniformly coupled system for $n=3$ and $n=4$ spins, utilising a generalization of the weak-coupling protocol already employed for 1- and 2-QST~\cite{Wojcik2005, Lorenzo2017b, Apollaro2015, Lorenzo2015}.

The paper is organised as follows: in Sec.~\ref{S.AF} we set the stage and derive expressions for the average fidelity and for the variance of fidelity of an arbitrary $n$-iQST protocol in terms of dynamical map elements. After having discussed in Sec. \ref{sectre} the special case of sequential transmission, which can be employed for $n$-QST only, in Sec.~\ref{S.nqst}, we derive the dynamical map elements for a quantum channel modeled by an $U(1)$-symmetric spin-$\frac{1}{2}$ Hamiltonian. We employ these results in Sec.~\ref{S.XX}, where we show that, for a quantum channel modeled by an $XX$ Hamiltonian, efficient 3- and 4-iQST is achievable via the weak-coupling protocol; in Sec. \ref{secconclu} we draw our conclusions. Finally, in the Appendix we report the explicit expressions for the elements of the dynamical maps used in the paper.

\section{Average fidelity and variance}\label{S.AF}

As stated in the Introduction, to assess the performance of an $n$-iQST protocol, the same quantum-information theoretical tools used for $n$-QST can be utilised. An important figure of merit for the efficiency of a QST protocol is the average fidelity $\average{F}$, defined as the fidelity averaged over all pure input states with respect to the unitarily invariant measure,
\begin{align}
\label{E_AvFid}
\average{F}=\frac{1}{\Omega}\int_{\Omega}d\Omega~F\left(\ket{\Psi},\rho(t)\right)~,
\end{align}
with $\Omega$ denoting the space of pure states and
\begin{align}
\label{E_UJfid}
F\left(\ket{\Psi},\rho(t)\right)=\bra{\Psi}\rho\ket{\Psi}
\end{align}
is the Uhlmann–Jozsa fidelity~\cite{Jozsa1994}. The evaluation of $\average{F}$ requires a parametrization of the pure state vector space in order to carry out integration. While this can be easily done for systems having low Hilbert space dimension (as it is the case for one~\cite{Bose2003} and two qubits~\cite{Lorenzo2015}), for an arbitrary pure state in $d$-dimensions, one needs to integrate over $2\left(d-1\right)$ reals parameters. Alternative methods have been devised, involving invariant integration over the $SU(d)$ group~\cite{Bagan2003,Bayat2007a}, or algebraic approaches using products of Pauli matrices~\cite{Cabrera2007}. However, in much of the existing literature about high-dimensional systems, the fidelity between quantum states has been used in a purely information-theoretical perspective, see, e.g., Ref.~\cite{Liang2018a} and references therein, and very little is known about the dynamics of the average fidelity in technologically relevant scenarios, such as the interacting many-body quantum state transfer protocols we are interested in.

Here, we develop an approach to the $n$ interacting qubits QST combining the formalism of quantum dynamical maps and invariant $SU(d)$ group integration, in order to obtain the average fidelity of an arbitrary $n$-qubit QST protocol.
To this end, consider the  map $\Lambda:\mathcal{S}\rightarrow\mathcal{R}$ that sends the input state of the sender into the output state of the receiver~\cite{Lorenzo:2021tnl}
\begin{align}\label{map1}
	\rhodm^R=\Lambda[\rhodm^S]~.
\end{align}
Taking an initial pure state for the sender $\rhodm^S = \ket{s}\!\!\bra{s}$, and expressing the map $\Lambda$ in its Kraus decomposition $\Lambda[\rhodm] = \sum_k\Ek\rhodm\Ekc$, with $\sum_k\Ekc\Ek = \mathbf{I}$, the output fidelity reads
\begin{align}\label{Ft2}
	F=\sum_k\bra{s}\Ek\ket{s}\!\!\bra{s}\Ekc\ket{s}=\sum_k\left|\bra{s}\Ek\ket{s}\right|^2~.
\end{align}
The average fidelity over all possible input states can be obtained by integrating Eq. \eqref{Ft2} with respect to the Haar measure on the unitary group $U_\mathcal{S}$ acting on $\mathcal{S}$ \cite{collins2006}:
\begin{align}\label{avFt1}
	\average{F}=\frac{1}{\Omega}\int_{\Omega}d\Omega~\sum_k\left|\bra{s}\Ek\ket{s}\right|^2=\sum_k\int dU_{\mathcal{S}} \left| \bra{\tilde{s}}\! \hat{U}^\dagger_\mathcal{S}\,\Ek^t\, \hat{U}_\mathcal{S}\!\ket{\tilde{s}}\right|^2~,
\end{align}
where we write the input state $\ket{s}$ as the unitarily transformed reference state $\ket{\tilde{s}}$.\\
Performing integration we arrive at 
\begin{align}\label{avFt2}
	\average{F} &= \sum_k\dfrac{1}{\dS(\dS+1)}\left({\rm Tr}\{\Ekc\Ek\}+{\rm Tr}\{\Ekc\}{\rm Tr}\{\Ek\}\right) \nonumber \\
	&= \dfrac{1}{(\dS+1)}+\dfrac{1}{\dS(\dS+1)}\sum_k\left|{\rm Tr}\{\Ek\}\right|^2~,
\end{align}
where $\dS$ stands for the dimension of the Hilbert space $\mathcal{S}$.
Note that the same expression can also be used to evaluate the average gate fidelity of a quantum channel~\cite{Nielsen2002, Johnston2011}.
Expressing the input state $\ket{s}$ in some complete orthonormal basis of the sender Hilbert space $\{\ketS{0},..\ketS{\dS{-}1}\}$,
\begin{align}
	\rhodm^R = \Lambda[\rhodm^S] = \sum_k\sum_{n,m=0}^{\dS-1} a_n a_m^* \,\,\Ek\ketbraS{n}{m}\Ekc~,
\end{align}
and choosing a basis also for the receiver Hilbert space $\mathcal{R}$, $\{\ketR{0},..\ketR{\dR{-}1}\}$, we  obtain
\begin{align}\label{E_map}
	\rhodm^R = \Lambda[\rhodm^S] &= \sum_k
	\sum_{n,m=0}^{\dS-1}
	\sum_{i,j=0}^{\dR-1}a_n a_m^* \,\, \braR{i}\Ek\ketbraS{n}{m}\Ekc\ketR{j} \,\,\ketbraR{i}{j} \nonumber \\
	&= \sum_{n,m=0}^{\dS-1}\sum_{i,j=0}^{\dR-1} A_{ij}^{nm}a_n a_m^* \ketbraR{i}{j}~,
\end{align}
and, hence, define the elements of the map $\Lambda$
\begin{align}\label{E_map_el}
	A_{ij}^{nm}=\sum_k \braR{i}\Ek\ketbraS{n}{m}\Ekc\ketR{j} \, .
\end{align}
The dynamical map in Eq.~\eqref{E_map} is usually represented as a $ij\times nm$ matrix acting on the input (sender) density matrix $\rho^S$ expressed as a column vector and giving the output (receiver) density matrix, $\vec{\rho}\,^R=A(t)\vec{\rho}\,^S$. In the following, we will consider only the case where the Hilbert spaces $\mathcal{S}$ and $\mathcal{R}$  have the same dimensions, i.e. $\dS = \dR = d$,  as our aim is to apply the present formalism to the $n$-iQST protocol.

The fidelity between $\rhodm^R$ and an arbitrary pure input state
\begin{align}
	\label{E_input}
	\ket{\Psi}=\sum_{p=0}^{d-1} a_p\ketS{p}~,
\end{align}
is given in terms of the dynamical map elements in Eq.~\eqref{E_map_el}, by
\begin{align}
	\label{E.Fid_dis}
F\left(\ket{\Psi},\rho\right)=\bra{\Psi}\rho\ket{\Psi}=\sum_{pqijnm=0}^{d-1}a_p^*a_qa_na_m^*A_{ij}^{nm}\delta_{pi}\delta_{jq}=\sum_{ijnm=0}^{d-1}a_i^*a_ja_na_m^*A_{ij}^{nm}~,
\end{align}
where all of the $a$'s refer to the initial state of Eq.~\eqref{E_input}, while the $\delta$'s arise from choosing the same basis for $\rho^S$ and $\rho^R$.
Although from Eq.~\eqref{E.Fid_dis} the full probability distribution function (PDF) for the fidelity can be derived for a given map, we will focus, in the following, only on the first and second moments of the distribution for arbitrary maps. Related results for the PDF between arbitrary quantum states can be found in Ref.~\cite{Zyczkowski2005}.

In order to evaluate the average fidelity, we use the results of Ref.~\cite{Bayat2007a}, where it is shown that the only non-zero averages are given by
\begin{align}
\label{E_averages}
\average{\left|a_i\right|^2}=\frac{1}{d}~,~\average{\left|a_i\right|^4}=\frac{2}{d\left(d+1\right)}~,~\average{\left|a_i\right|^2\left|a_j\right|^2}_{i\neq j}=\frac{1}{d\left(d+1\right)}~,
\end{align}
which, in our case, yield
\begin{align}
\label{E_AF}
\average{F}=\frac{1}{d\left(d+1\right)}\left(2\sum_{i=0}^{d-1}A_{ii}^{ii}+\sum_{i\neq j=0}^{d-1}A_{ii}^{jj}+2\Re\left\{\sum_{i>j=0}^{d-1}A_{ij}^{ij}\right\}\right)~.
\end{align}
Similarly to the results of Refs.~\cite{Nielsen2002,Johnston2011,Pedersen2008,Pedersen2007,Mayer2018}, Eq.~\eqref{E_AF} provides quite a simple expression for the average fidelity in terms of the quantum dynamical map elements (notice, indeed, that only $\frac{d}{2}\left(3d-1\right)$, out of $d^4$, of the quantum map's elements really matter). It will be exploited many times in the following, in order to assess the performance of $d$-dimensional QST in various physically relevant cases. 

It is straightforward to show that Eq.~\eqref{E_AF} encompasses both the trivial case $\Phi(t)=\mathbf{I}$, i.e., $A_{ij}^{nm}=\delta_{in}\delta_{jm}$, yielding $\average{F}=1$, and the LOCC-limit. The latter, in particular, is obtained from the first two contributions to $\average F$, which come from those elements of the map $A$ connecting all of the diagonal elements of the density matrices of $S$ and $R$. The third term, instead, is due to off-diagonal map elements, connecting input to output coherences. It has, thus, a purely quantum origin, and it disappears for a classical map. The LOCC limit is obtained by setting $A_{ij}^{nm}=\delta_{ij}\delta_{nm}$, yielding $\average{F}=\frac{2}{d+1}$. This is the maximum possible value achievable from the first two terms only, as one can infer by using the following constraints on the map elements, which are obtained from the fact that $A$ represents a CPTP map in the chosen basis:
\begin{align}
	\sum_{i}A_{ii}^{nm} &= \sum_{i}\sum_k\braR{i}\Ek\ketbraS{n}{m}\Ekc\ketR{i} \nonumber \\
	&= \sum_{i}\sum_k\braS{m}\Ekc\ketbraR{i}{i}\Ek\ketS{n} \nonumber \\
	&= \braS{m}\sum_k \Ekc\Ek\ketS{n} \nonumber \\
	&= \delta_{nm}~,
\end{align}
\begin{align}
	\sum_{i}{A_{ii}^{nn}}=1\rightarrow 0\leq A_{ii}^{nn} \leq 1~, \qquad 	\sum_{inm}{A_{ii}^{nm}}=d~,
\end{align}
\begin{align}
	A_{ij}^{nm}=\left(A_{mn}^{ji}\right)^*~, \qquad A_{ij}^{ij}=\left(A_{ji}^{ji}\right)^*~.
\end{align}
Thus, the last quantum term is crucial for a good performance of the channel, and it becomes more and more relevant by increasing the dimension $d$, as it may amount to a value up to $1-\frac{2}{d-1}=\frac{d-1}{d+1}$.
As a final comment to the expression above for $\average F$ in Eq.~\eqref{E_AF}, we notice that map's elements connecting different off-diagonal elements of $\rho^R$ and $\rho^S$ do not play any role, as they are averaged out by the integral over unitaries.

\subsection{Variance of the Fidelity}
Up to now, we have focused on the average fidelity. However, this does not give a complete characterization of the performance of a quantum state transfer protocol, and thus, we shall now turn our attention to the variance of the fidelity distribution, which provides a description of the dispersion of the values of $F$ for different input states. To obtain the variance, which is defined in the usual way as
\begin{align}
	\label{E_var_1}
	\left(\Delta F\right)^2=\average{F^2}-\average{F}^2~,
\end{align}
we need to evaluate the second moment of the distribution of the possible values of the fidelity taken over all pure input states, Eq.~\eqref{E.Fid_dis}, which 
is obtained by averaging the following expression
\begin{align}
	F^2\left(\ket{\Psi},\rho\right)=\sum_{ijnmpqrs=0}^{d-1}a_i^*a_ja_na_m^*a_p^*a_qa_ra_s^*A_{ij}^{nm}A_{pq}^{rs}~.
\end{align}
Using, again, the results of Ref.~\cite{Mello1990}, we get


\begin{align}
	\average{F^2} &= \frac{1}{d (d( + 1)(d + 2)(d + 3)} \times \nonumber \\
	\sum_{i, m, p, s = 0}^{d-1} & \left(
	(A_{ii}^{mm}+A_{im}^{im})(A_{pp}^{ss}+A_{ps}^{ps}) +
	(A_{ii}^{pm}+A_{ip}^{im})(A_{pm}^{ss}+A_{ps}^{ms}) \ + \right. \nonumber \\[-1ex]
	& \ \ (A_{ii}^{sm}+A_{is}^{im})(A_{pm}^{ps}+A_{pp}^{ms}) + (A_{im}^{pm}+A_{ip}^{mm})(A_{pi}^{ss}+A_{ps}^{is}) \ + \nonumber \\[1ex] & \ \left. (A_{im}^{sm}+A_{is}^{mm})(A_{pi}^{ps}+A_{pp}^{is}) + (A_{ip}^{sm}+A_{is}^{pm})(A_{pi}^{ms}+A_{pm}^{is}) \right) ~.
	\label{E_f2}
\end{align}
As an easy check for this expression, if the map is the unit map, i.e., $A_{ij}^{nm}=\delta_{in}\delta_{jm}$, then 	$\average{F^2}=1$. 
In the next sections, we will explicitly evaluate the second moment and the variance for the case of an $n$-qubit channel, both in the absence and in the presence of interactions.

\section{\textit{n}-QST over independent channels}
\label{sectre}
\begin{figure}[!ht]
    \centering
	\includegraphics[width=0.7\textwidth]{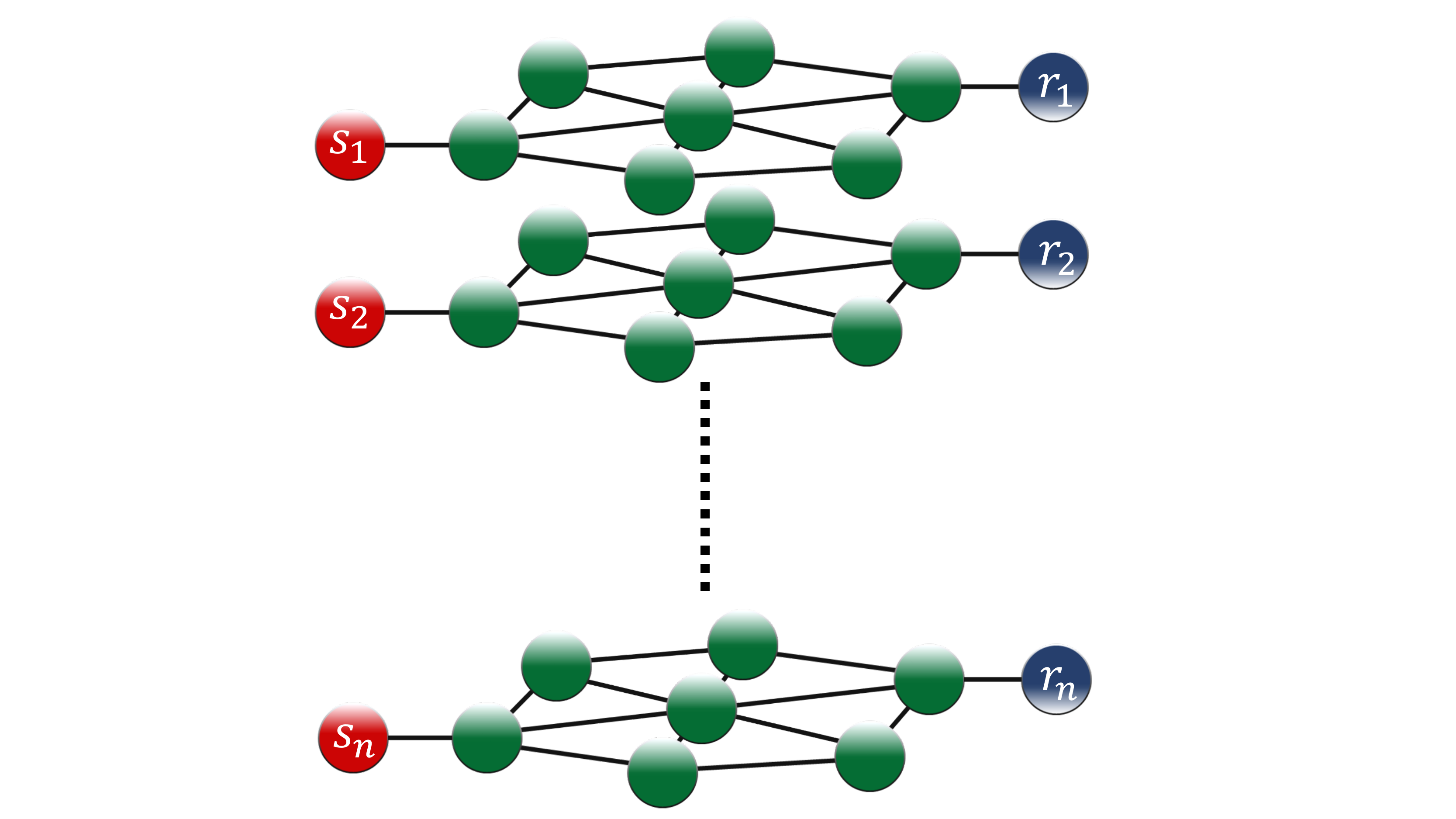}
	\caption{Sketch of the parallel protocol for the transfer of $n$ non-interacting qubits via $n$ independent channels.}
	\label{F.Figure_ind}
\end{figure}

Before turning our attention to the transfer of $n$ interacting qubits via a single quantum channel, embodied by a spin-$\frac{1}{2}$ chain, let us analyse the case of $n$ non-interacting qubits transferred `in parallel' across $n$-independent channels. The topology of the independent channels can be arbitrary, and here we model them as $U(1)$-symmetric spin-$\frac{1}{2}$ networks as depicted in Fig.~\ref{F.Figure_ind}. Besides being interesting in itself, this case will serve as a benchmark to compare the performance of other, more involved, transmission set-ups.

With this approach, we need to transfer one qubit per channel; thus, we can make use of the well known result that, for an initially fully polarized channel+receiver system, 1-QST is completely determined by a single parameter, the transition amplitude for one-spin excitation to be transferred from sender to receiver~\cite{Bose2003}. Such a transition amplitude, that can be taken to be real (see also next section), and that we call here $f$, can be manipulated in various ways, using any of the control schemes reported in the literature. Whatever approach one employs to maximize $f$, once its value is set, the dynamical map representing the single qubit transmission has the amplitude damping form, and it is reported in Eq.~\eqref{E_1ex_map} of \ref{appendix1}.

\begin{figure}[!ht]
    \centering
	\includegraphics[width =0.49\textwidth]{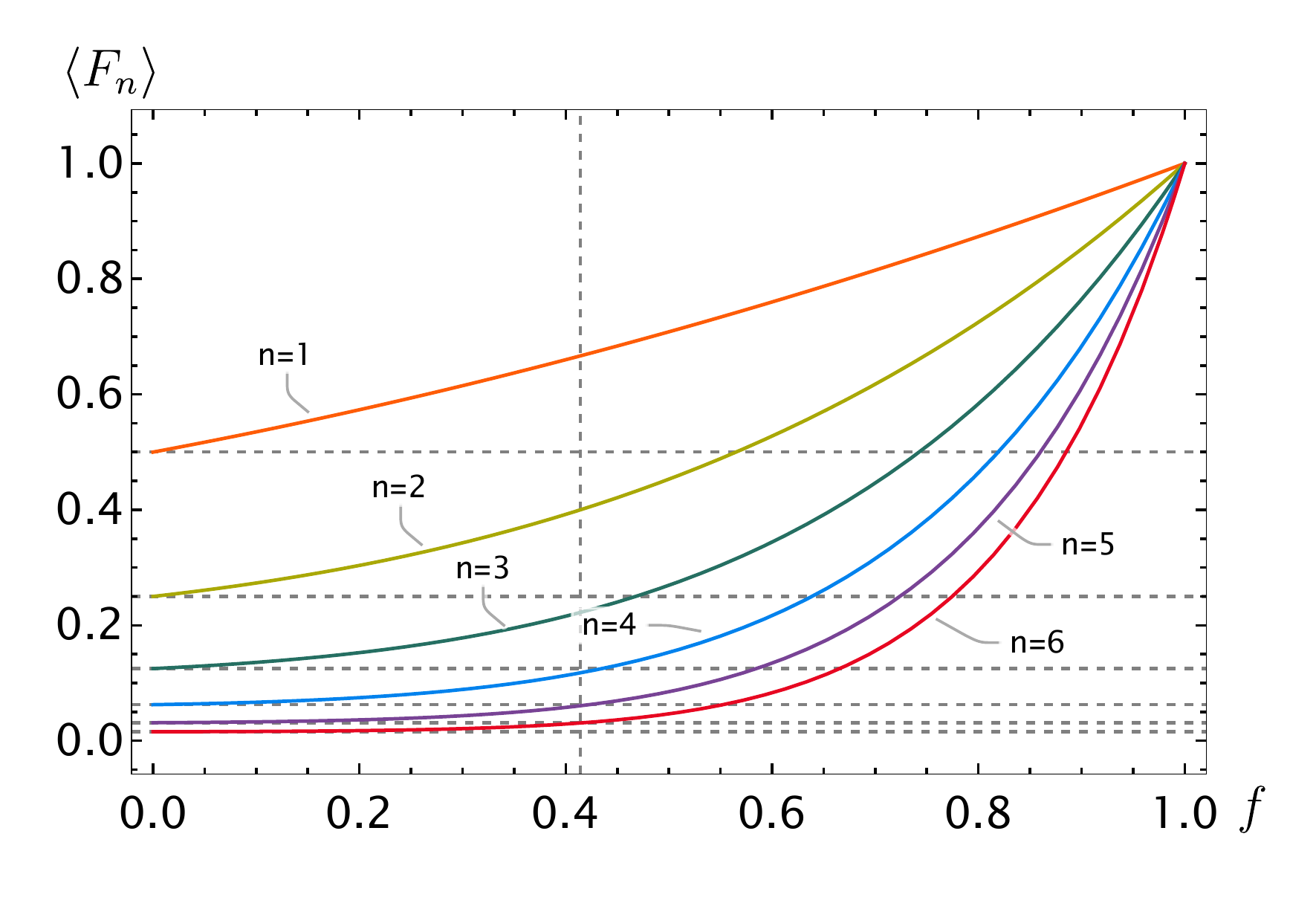}
	\includegraphics[width =0.49\textwidth]{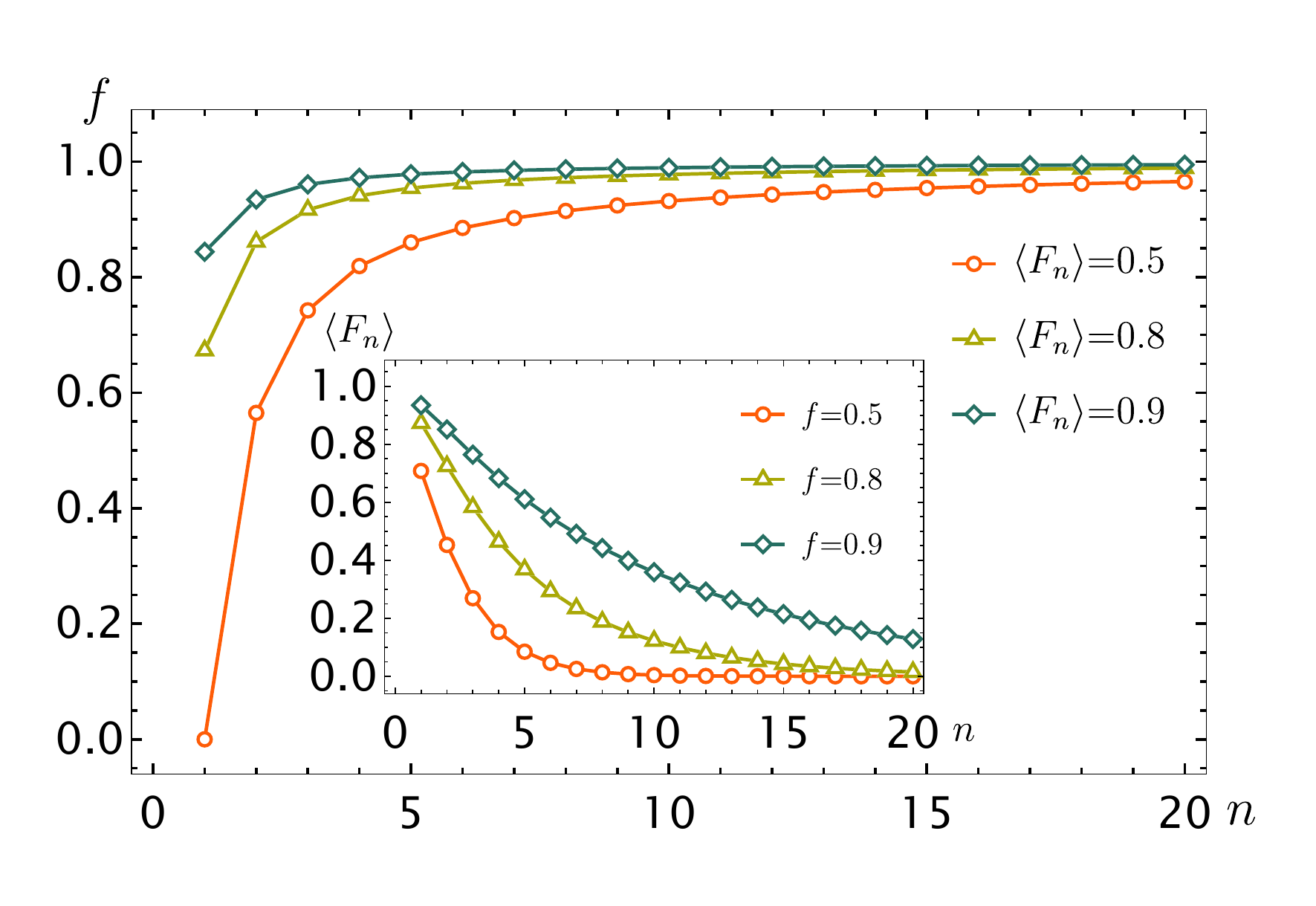}
	\caption{(left) $n$-QST average fidelity $\average{F_n}$ vs. transition amplitude $f$ for $n=1,2,3,4,5,6$ (top-down). Horizontal dotted lines represent the LOCC limit and the vertical dotted line is set at $f=\sqrt{2}-1$. It is possible to appreciate that the higher the dimensionality of the state to be transferred, the closer to $1$ the transition amplitude $f$ has to be in order to achieve a high-quality QST. (right) Required transition amplitude $f$ as a function of the number of qubits $n$ in order to achieve $\bar{F}_n=0.5,0.8,0.9$ (main); achieved average $n$-QST fidelity vs. $n$ for single-particle transition amplitudes $f=0.5,0.8,0.9$ (inset).}
	\label{FLOCC_fid}
\end{figure}

Under the assumption that the dynamical maps $A$ of the various parallel transmission channels are all the same, the $n$-QST average fidelity of Eq.~\eqref{E_averages} becomes
\begin{align}
	\label{E_F_ind}
	\average{F_n}=\frac{1}{d+1}+\frac{1}{d\left(d+1\right)}\left|1+ f\right|^{2n}~.
\end{align}
In order to overcome the classical $n$-QST LOCC limit, the single-particle amplitude $f$ has to exceed the 1-QST LOCC limit, $f>\sqrt{2}-1$. This is independent of the number of qubits $n$ and results in a polynomially decaying fidelity for decreasing $f$. In Fig.~\ref{FLOCC_fid} we report the  $n$-QST average fidelity as a function of the transition amplitude $f$ and show that for a large number of qubits, e.g., $n\simeq 20$, high-quality QST is achieved only for almost-unit single-particle transition amplitude.

From Eq. \eqref{E_F_ind}, we notice that the average fidelity $\average{F_n} \neq \prod_{i=1}^{n}\average{F_1}$. In fact, the product of $\average{F_1}$ gives the average fidelity of the QST only for fully factorized states, i.e., if $\ket{\Psi}_n=\bigotimes_{i=1}^{n}\ket{\psi}_i$. As, in general, for independent processes, the average of the product of a set of random variables is equal to the product of their averages, we conclude that, when $n>2$, entanglement gives rise to a  breakdown of independence in the parallel transfer processes, and it does so in such a way as to reduce the fidelity. 

In order to quantify the effect of entanglement on the $n$-QST with the set-up reported in Fig.~\ref{F.Figure_ind}, we introduce, as a figure of merit, the ratio $R$ between the fidelity of the $n$-QST of the subset of product states and of the full set of states. $R$ can then be expressed both as a function of the transition amplitude $f$ and as a function of the fidelity. 

\begin{figure}[!ht]
\centering
\includegraphics[width =0.49\textwidth]{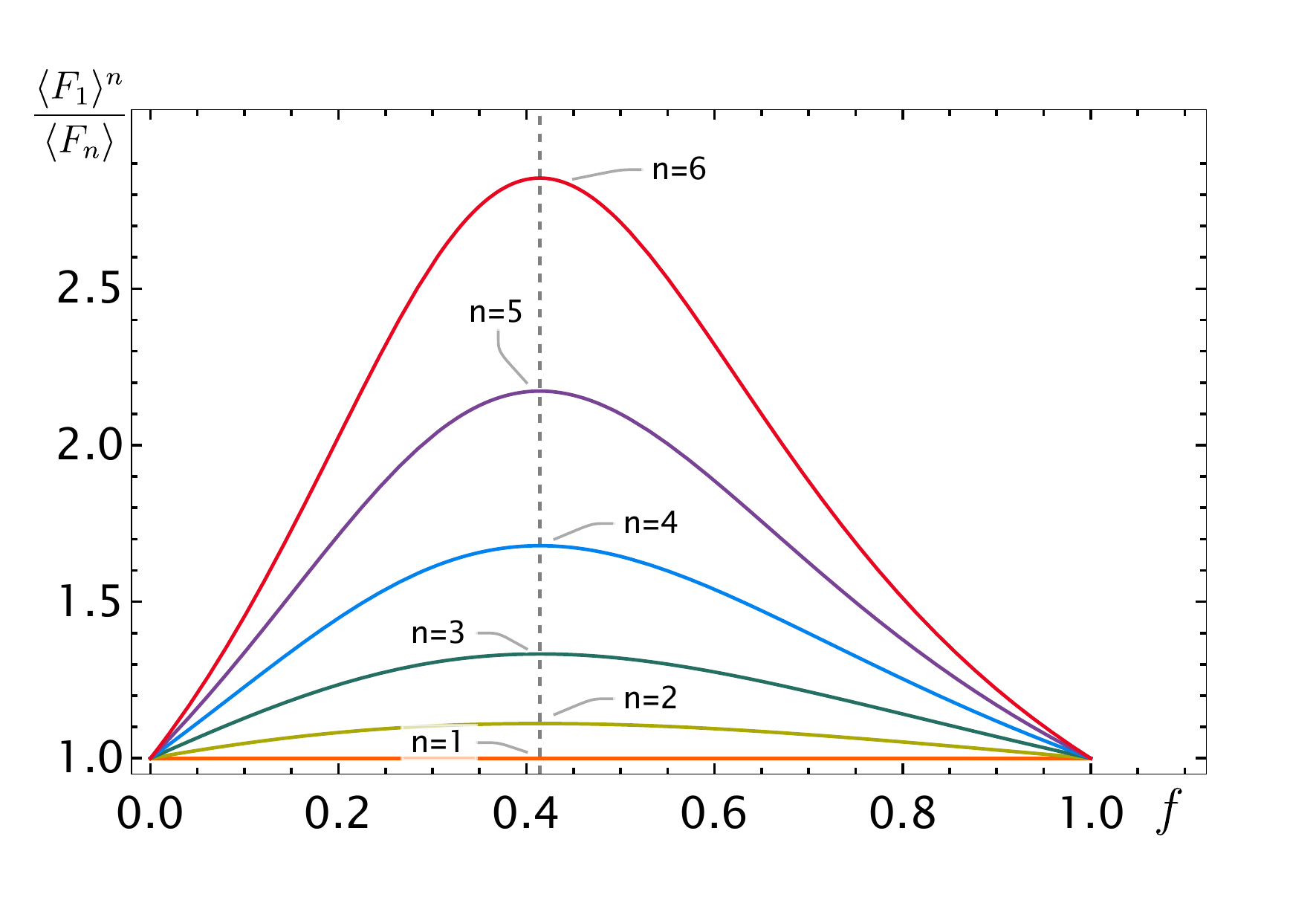}
\includegraphics[width=0.49\textwidth]{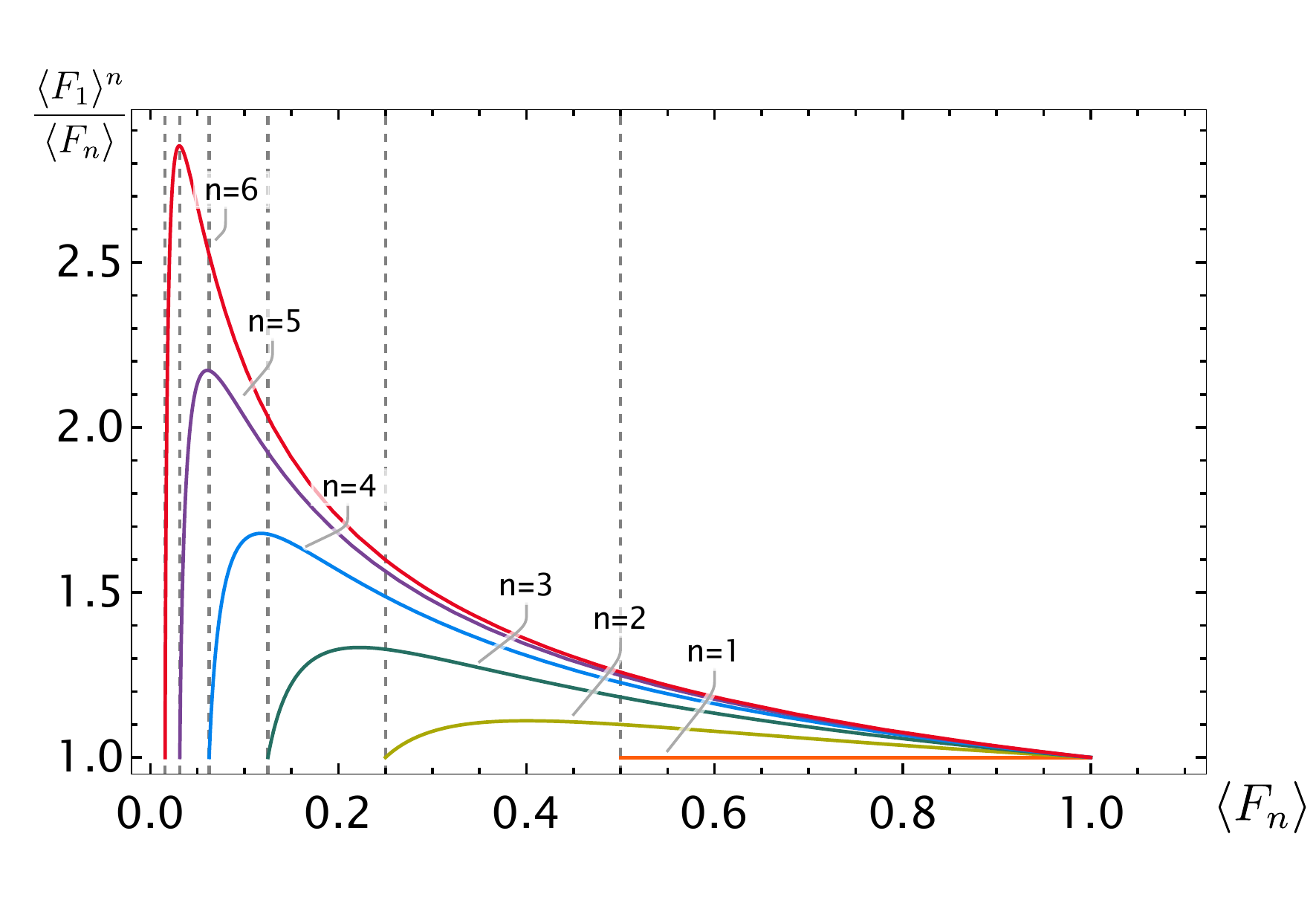}
\caption{(left) Plot of $R(f)=\frac{\average{F_1}^n}{\average{F_n}}$ in Eq.~\eqref{E_prod_ent_f}. Here, the black, dashed line, signals the maximum of the ratio $R(f)$, occurring at $f_{LOCC}=\sqrt{2}-1$, corresponding to the LOCC limit for $\average{F_n}$. Curves are for $n=1,2,\dots,6$ (bottom-up). For $n>1$ the  $n$-QST fidelity of the subset of product states, $\average{F_1}^n$, is always greater than that of the full set of states $\average{F_n}$, at fixed transition amplitude $f\neq 0,1$. (right)  Plot of $R(F)=\frac{\average{F_1}^n}{\average{F_n}}$ in Eq.~\eqref{E_prod_ent_F}, as a function of $F\equiv \average{F_n}$. Here, the black, dashed lines, signal the maximum of the ratio $R(F)$ occurring at $F_{LOCC}=\frac{2}{d+1}$, corresponding to the LOCC limit for $\average{F_n}$. Curves are for $n=1,2,\dots,6$ (bottom-up). For $n>1$ the $n$-QST fidelity of the subset of product states, $\average{F_1}^n$, is always greater than that of the full set of states $\average{F_n}$, at fixed average fidelity of the latter $\average{F_n}\neq0,1$.}
\label{F.prod_ent_f}
\end{figure}

In the first case, we obtain
\begin{align}
	\label{E_prod_ent_f}
	R(f)=\frac{\average{F_1}^n}{\average{F_n}}=\frac{3^{-n}\left(2^n+1\right)(f(f+2)+3)^n}{(f+1)^{2n}+2^n}\geq 1 \,,
\end{align}
with equality holding for $f = 0, 1$. The maximum of $R(f)$ occurs  precisely at the amplitude value $f_{LOCC}=\sqrt{2}-1$ that saturates the LOCC-limit, yielding $R(f_{LOCC})=\frac{1}{2} \left(\left(\frac{2}{3}\right)^n+\left(\frac{4}{3}\right)^n-2\right)$. 
Eq.~\eqref{E_prod_ent_f} is displayed in the left panel of Fig.~\ref{F.prod_ent_f}, which shows that, at fixed transition amplitude $f$, product states enjoy a higher $n$-QST fidelity than entangled states. To further sustain the claim that product states are better transferred than entangled ones by independent channels, it is instructive to report also the ratio, $R(F)$, between the $n$-QST fidelity averaged over product states only, and over the full set of states at a given value, $F$, of the average fidelity. From Eq.~\eqref{E_F_ind}, we obtain that $\average{F_n}=F$ when the transition amplitude attains the value $f_F=\sqrt{2}\left(\left(2^n+1\right) \left(F-\frac{1}{2^n+1}\right)\right)^{\frac{1}{2n}}-1$. Using this value of the transition amplitude, $f_F$, into the $n$-QST average fidelity over the subset of product states $\average{F_1}^n$, we get the ratio $R$ to be
\begin{align}
	\label{E_prod_ent_F}
	R(F)=\frac{ \left(1+\left(2^n F +F-1\right)^{\frac{1}{n}}\right)^n}{3^{n}F}\xrightarrow{n\rightarrow \infty} F^{-\frac{1}{3}}
\end{align}
In the right panel of Fig.~\ref{F.prod_ent_f}, we plot the ratio in Eq.~\eqref{E_prod_ent_F} and show that, for $n>1$, at a fixed fidelity of the full set of states, the subset of product states enjoys a higher $n$-QST average fidelity.

\begin{figure}[!ht]
    \centering
	\includegraphics[width=0.49\textwidth]{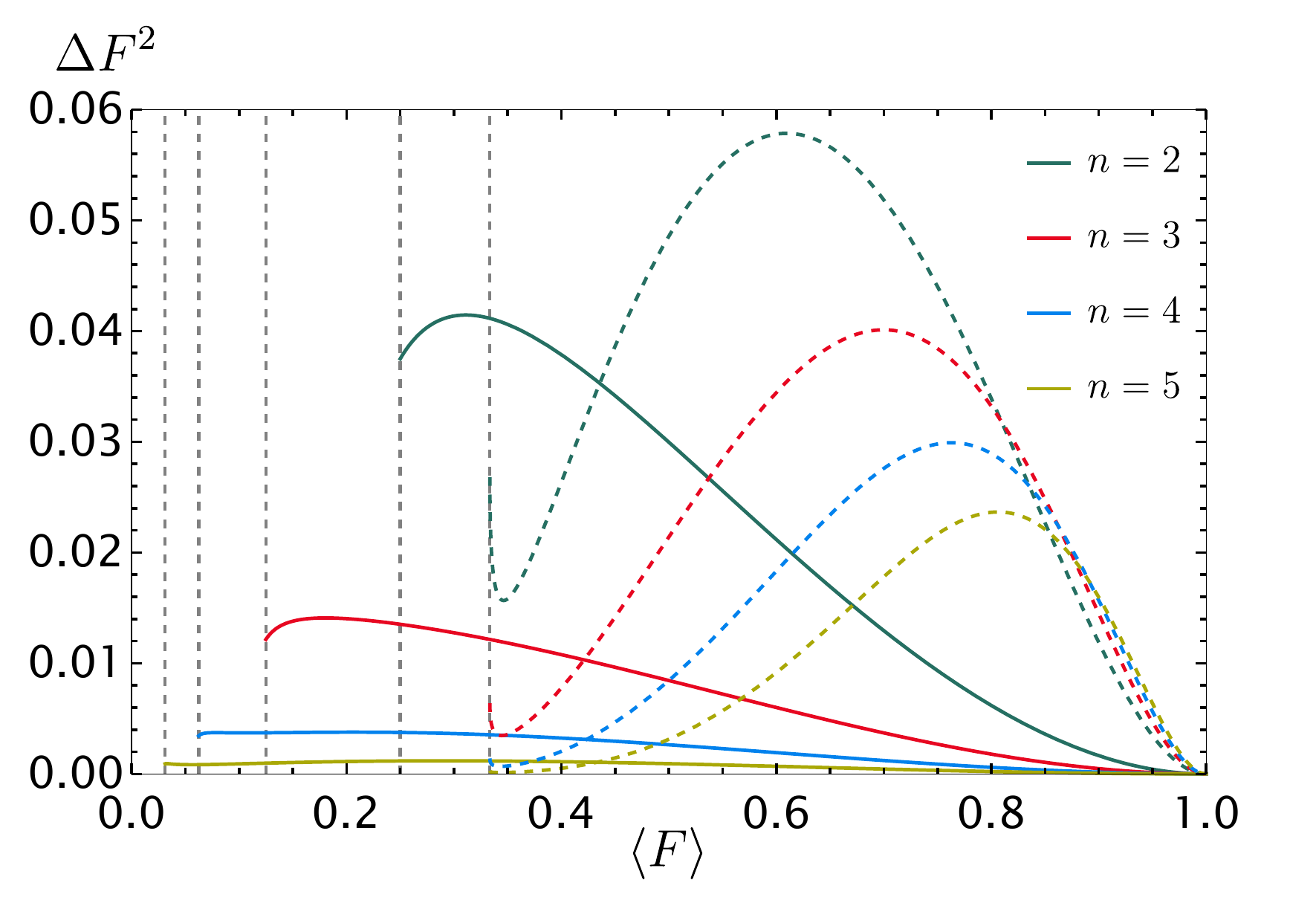}
    \includegraphics[width =0.49\textwidth]{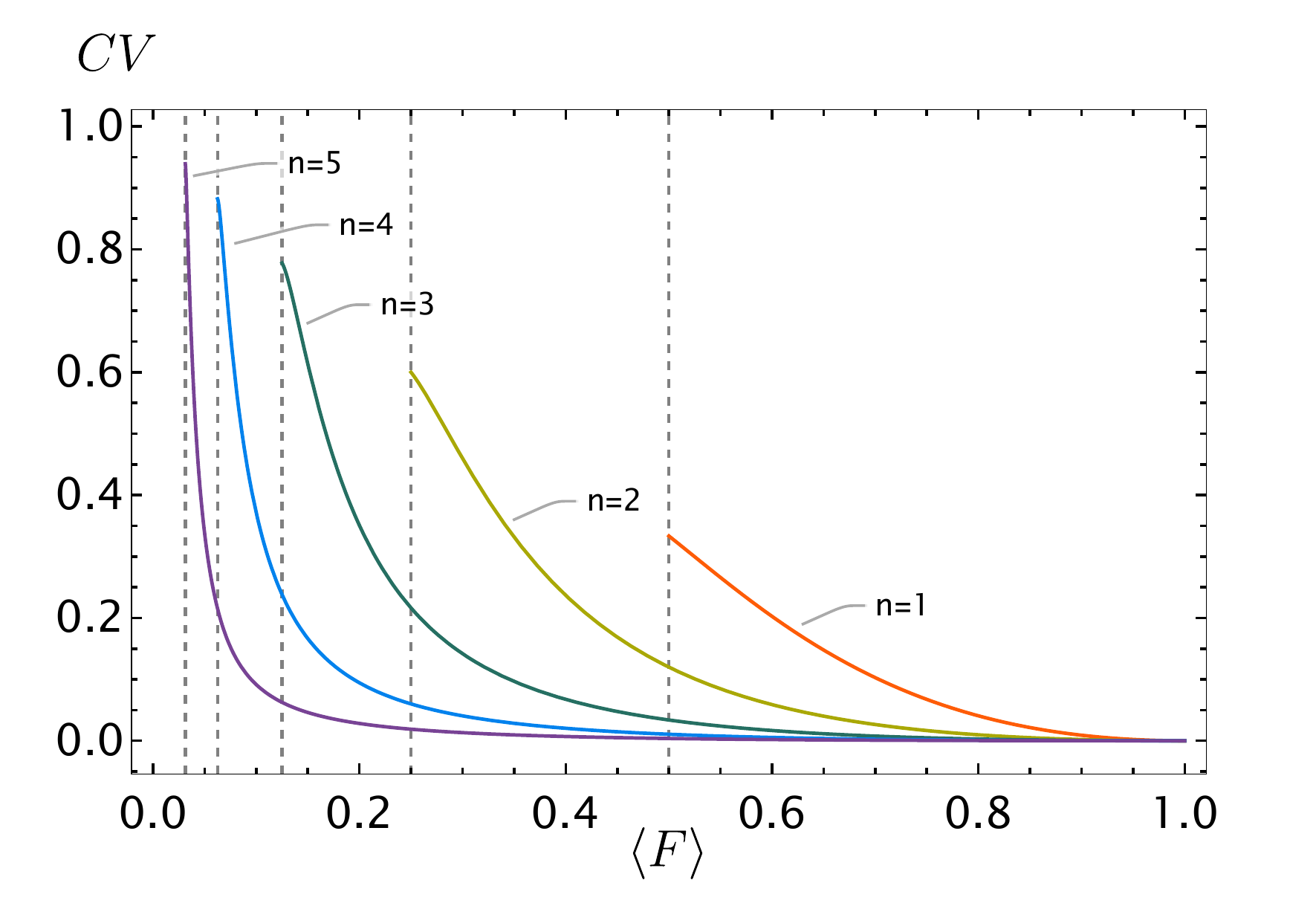}
\caption{(left) Variance $\Delta F^2$ vs average fidelity $\average{F}$ for $n$-QST across independent channels for product states (dotted lines) and arbitrary states (continuous lines). Curves are for $n=2,3,4,
5$, respectively red, blue, black, and green. (right) Coefficient of variation vs. $\average{F_n}$ for $n=1,2,\dots,5$, as given in Eq.~\eqref{E_kz}. Black, dashed lines correspond to the random guess scenario, i.e., $f=0$ yielding $\average{F_n}=\frac{1}{d}$.}
\label{F.prod_ent_s}
\end{figure}

Let us now turn our attention to the variance evaluated for the case of independent channels. The variance for an $n$-qubit arbitrary state is given by Eq.~\eqref{E_var_1}, whereas the variance restricted to product states input follows the law $\left(\Delta F\right)^2=\average{F_1^2}^n-\average{F_1}^{2n}$, where $F_1$ is the 1-QST fidelity. From the left panel in Fig.~\ref{F.prod_ent_s} we see that the variance for product states is greater for high-average fidelity values. Intuitively, this can be explained by noticing that the set of separable pure states is of zero measure within the set of pure states. Hence, there are less states in the neighborhood of the sender state giving the targeted fidelity, resulting in a flatter probability distribution function of the fidelity, and an increased variance with respect to the variance obtained for the full set of pure states.  

To conclude this section, we report the coefficient of variation $CV$ as a figure of merit of the relative variability for the fidelity
\begin{align}
	\label{E_kz}
	CV=\frac{\sigma}{\average{F_n}}
	=\sqrt{\frac{\average{F_n^2}}{\average{F_n}^2}-1}~,
\end{align}
where $\sigma=\Delta F$ is the standard deviation. For the case of independent channels we are dealing with here,
$CV$ is reported in the right panel of Fig.~\ref{F.prod_ent_s} as a function of $\average{F_n}$. From the plot, we can conclude that, as perhaps expected, the relative dispersion becomes smaller as the average fidelity increases. Moreover, at fixed fidelity, $CV$ tends to zero with increasing $n$, meaning that the average fidelity $\average{F_n}$ is a self-averaging quantity, for $n\gg 1$ and $\average{F_n}>\frac{1}{d}$.


\section{\textit{n}-iQST via spin chains}\label{S.nqst}
In this section, we discuss quantum transfer for $n$ interacting qubits via spin chains, where the sender's, the receiver's as well as the wire's qubit dynamics are described by the same Hamiltonian.
A much investigated QST protocol models the channel as an open boundary linear spin-$\frac{1}{2}$ chain~\cite{Bose2007}, where the sender and the receiver are located at opposite edges, see Fig.~\ref{F.Figure} for a pictorial representation of the protocol applied to $n$-qubit QST via weak-coupling. This protocol has been shown to allow for both one- and two-qubit high-fidelity QST under a variety of different dynamical parameters~\cite{Wojcik2005,Banchi2011b,Hermes2020a,Almeida2015,Pavlis2016a}. For $n>2$, however, only a few results have been obtained up to now.

Here, after revisiting the 1- and 2-QST using our formalism, we will move to $n>2$ iQST, and propose a protocol for high-quality 3- and 4-iQST.
In particular, we will analyse the set-up where the whole sender+wire+receiver system has U(1) symmetry, i.e., it preserves the total magnetization along the $z$-axis, and we assume both our channel (i.e., the wire), and the receiver qubits to be initially fully polarized. This class of $U(1)$-symmetric Hamiltonians encompasses a large number of models, including, e.g., the following general Heisenberg-type Hamiltonian
\begin{align}
\label{E_Ham_Hei}
H=\sum_{n}\sum_r\left( J_{n,r}\left(\hat{\sigma}^x_{n}\hat{\sigma}^x_{n+r}+\hat{\sigma}^y_{n}\hat{\sigma}^y_{n+r}\right)+\Delta_{n,r}\hat{\sigma}^z_{n}\hat{\sigma}^z_{n+r}+h_n\hat{\sigma}^z_{n}\right)~,
\end{align}
where the sum runs over lattice sites $n$ and  interaction range $r$, and where we allowed for an anisotropic exchange coupling between spins along the $z$-axis (so that, in general, $J_{n,r} \ne \Delta_{n,r}$), and for an external transverse field $h_n$.


An arbitrary initial pure state, having up to $n_s$ excitations, i.e., spin-up states $\ket{1}$, located in the sender block, with the rest of the system being in $\ket{\mathbf{0}}\equiv\ket{00\dots00}$, can be written in the computational basis (with  $\ket{n}\equiv\ket{0_1, 0_2, \dots 1_n\dots 0}$), as
\begin{align}
\label{E_sender}
\ket{\Psi(0)} = \ &a_0 \ket{\mathbf{0}} + \sum_{n\in\{n_s\}}a_n\ket{n} + \sum_{n<m\in \{n_s\}}a_{nm}\ket{nm} \ + \nonumber \\ &\sum_{n<m<p\in\{n_s\}}a_{nmp}\ket{nmp} + \sum_{n<m<p<q\in\{n_s\}}a_{nmpq}\ket{nmpq} + \dots
\end{align}
According to the dynamics generated by the Hamiltonian in Eq.~\eqref{E_Ham_Hei}, such an initial state evolves into~\cite{Apollaro2015}
\begin{align}
\label{E_evo}
\ket{\Psi(t)} = \ &a_0 \ket{\mathbf{0}} + \sum_{m=1}^N\left(\sum_{n\in \{n_s\}}a_n f_n^m\right)\ket{m}+\sum_{p<q=1}^N\left(\sum_{n<m\in \{n_s\}}a_{nm}f_{nm}^{pq}\right)\ket{pq} \ + \nonumber \\ &\sum_{r<s<t=1}^N \left(\sum_{n<m<p\in \{n_s\}}a_{nmp}f_{nmp}^{rst}\right)\ket{rst} \ + \nonumber \\
&\sum_{r<s<t<u=1}^N \left(\sum_{n<m<p<q\in \{n_s\}}a_{nmpq}f_{nmpq}^{rstu}\right)\ket{rstu} \ + \ \dots
\end{align}
where $f_{s_1s_2\ldots s_n}^{r_1r_2 \ldots r_n}=\bra{r_1r_2 \ldots r_n}e^{-i \hat{H}t}\ket{s_1s_2\ldots s_n}$ is the transition amplitude for $n$-excitation states from sites $s_1s_2\ldots s_n$ to $r_1r_2 \ldots r_n$.

The state of the receiver's qubits at sites $\{n_r\}$ is generally given by a density matrix, which can be obtained, via a lengthy but straightforward calculation, by tracing out all but the receiver's qubits. Finally, the dynamical map $A$ is derived from the latter by comparison with Eq.~\eqref{E_map}. Below, we give a brief overview of this procedure for the cases of $n=1$ and $n=2$, and then move to the 3- and 4-iQST.

\subsection{1-QST}\label{sS.1}
For the QST of 1 qubit, the average fidelity of Eq.~\eqref{E_AF} reads
\begin{align}
\label{E_AF1}
\average{F_1}=\frac{1}{6}\left(2\left(A_{00}^{00}+A_{11}^{11}\right)+A_{00}^{11}+A_{11}^{00}+2\Re\left\{A_{10}^{10}\right\}\right)~.
\end{align}
The non-zero map elements entering this expression (see \ref{appendix1}) are
\begin{align}
\label{E_1map}
	A_{00}^{00}=1~,~A_{11}^{11}=\left|f_{1}^{N}\right|^2~,~A_{10}^{10}=\left(f_{1}^{N}\right)^*~,~A_{00}^{11}=1-\left|f_{1}^{N}\right|^2~,
\end{align}
where $f_{1}^{N}=\bra{N}e^{-i \hat{H}t}\ket{1}$ is the transition amplitude of one-spin excitation to travel from the sender location $1$ to the receiver location $N$.
With these map elements, Eq.~\eqref{E_AF1} becomes
\begin{align}
\label{E_1QST_XX}
\average{F_1}=\frac{1}{2}+\frac{\left|f_1^N\right|^2}{6}+\frac{\Re\left\{f_1^N\right\}}{3}=\frac{1}{2}+\frac{\left|f_1^N\right|^2}{6}+\frac{\left|f_1^N\right| \cos\phi}{3}~,
\end{align}
which coincides with the result in Ref.~\cite{Bose2003}, and has been used also before, in Sec. \ref{sectre}, when analysing transmission along parallel channels. In the previous expression, $\phi$ is the argument of the complex number $f_1^N$. In order to maximise $\average{F_1}$, one chooses to perform a rotation on the receiver (or to apply a magnetic field to the quantum channel) such that $\cos \phi=1$, so that only the modulus of the transition amplitude matters.

\subsection{2-QST}\label{sS_2qst}

The transfer of a (possibly entangled) two-qubit state along a spin chain has been analysed, e.g. in Ref.~\cite{Apollaro2015}.  There, it was shown that the receiver's state and the fidelity can be written in terms of one- and two-spin excitation transfer amplitudes. The map allowing to obtain the output (receiver) state from the input (sender) one, is reported in Eq.~\eqref{E_2map} in \ref{appendix2}. Using these results, a lengthy but straightforward calculation gives for the average fidelity of  Eq.~\eqref{E_AF}
\begin{align}
\label{E_AF2}
\average{F_2}=&\frac{1}{4}+\frac{1}{20}\left(\left|f_1^N\right|^2+\left|f_2^{N-1}\right|^2+\left|f_{12}^{NN-1}\right|^2\right)\nonumber\\
&+\frac{1}{10}\Re\left\{f_1^N+f_2^{N-1}+f_{12}^{NN-1}+f_2^{N-1}\left(f_1^N\right)^*+f_{12}^{NN-1}\left(f_1^N\right)^*+f_{12}^{NN-1}\left(f_2^{N-1}\right)^*\right\}~,
\end{align}
which simplifies the expression already obtained in Ref.~\cite{Lorenzo2015}, and has a straightforward physical interpretation: to achieve unit fidelity, all of the excitations initially located on sender sites $i$ need to reach their mirror-symmetric sites $N+1-i$ located in the receiver's block, at the same time, with unit transition amplitude. Notice that a similar map has been used in Ref.~\cite{Apollaro2010a} to investigate the non-Markovian dynamics of two qubits coupled to spin environments. 

Let us stress here that, at variance with the 1-QST in Sec.~\ref{sS.1}, we can not, in general, turn the expression contained in the last term into a sum of transition amplitude moduli with a common phase, as the different transition amplitudes will have, in general, different arguments. The same will be true for the average fidelity of every $n>1$-QST protocol.

\subsection{n-iQST}
The $n$-iQST average fidelity $\average{F_n}$ can be derived, for a quantum map given by $U(1)$-symmetric Hamiltonians where the receiver and the channel are initially fully polarized, by a procedure similar to that outlined in Sec.~\ref{sS_2qst}. The $n$-qubit dynamical map (which we do not report for the sake of brevity) is derived in terms of $p\rightarrow q$ qubit transition amplitudes, with $p,q\in\left[1,n\right]$. A straightforward calculation yields
\begin{align}
\label{E_Fn}
\average{F_n}=\frac{1}{d}+\frac{1}{d\left(d+1\right)}\sum_{S}\left|f_S^S\right|^2+\frac{2}{d\left(d+1\right)}\Re\left\{\frac{1}{2}\sum_S f_S^S\left(1+\sum_{S'\neq S} f_{S'}^{S'}\right)^*\right\}~.
\end{align}
We recognise Eq.~\eqref{E_Fn} to have the same structure as $\average{F_2}$ in Eq.~\eqref{E_AF2}, with the classical contribution given by the sum of all transition probabilities between the mirror-symmetric partitions of the sender and receiver blocks, and the quantum part given by the product of all of the $n$-qubit transition amplitudes times the (complex conjugate) of all of the $m$-qubit transition amplitudes between the $S$ and $R$, with $m<n$. Before analysing this expression for the particular cases of $n=3$ and $n=4$, some observations are in order: The first term in Eq.~\eqref{E_Fn} corresponds to a random guess $\average{F_n}=\frac{1}{d}$; on the other hand, the summation in the second term runs over all possible transitions $i\in S\rightarrow i'\in R$ with $i=i'$, achieving a total number of $\sum_{r=1}^n\binom{n}{r}=2^n-1$. The sum of the first two terms, then, gives the LOCC limit $\average{F_n}=\frac{2}{d+1}$ if all transition probabilities are equal to unity. As mentioned before, the third terms contains the product of all the transition amplitudes that build up the coherence in the output state, and for this reason we consider this to be a purely quantum contribution.

Due to mirror-symmetry, we can employ a smarter notation in which $S=\{(i),(ij),(ijk),(ijkl),\dots\}$ are the labels of the sender sites (for the different $n$ values) and in which we label the receiver site $r=N+1-s$ by the same numbering, $s$, as its mirror-symmetric counterpart on the sender block. With this convention, Eq.~\eqref{E_Fn} can be rewritten in a more compact form as
\begin{align}
	\label{E_Fn_1}
	\average{F_n}=\frac{1}{d+1}+\frac{1}{d\left(d+1\right)}\left|1+\sum_S f_S^S\right|^2~.
\end{align}

\section{Efficient 3- and 4-iQST in the XX spin-\texorpdfstring{$\frac{1}{2}$} model with weak links}\label{S.XX}

\begin{figure}[!ht]
    \centering
	\includegraphics[width =0.8\textwidth]{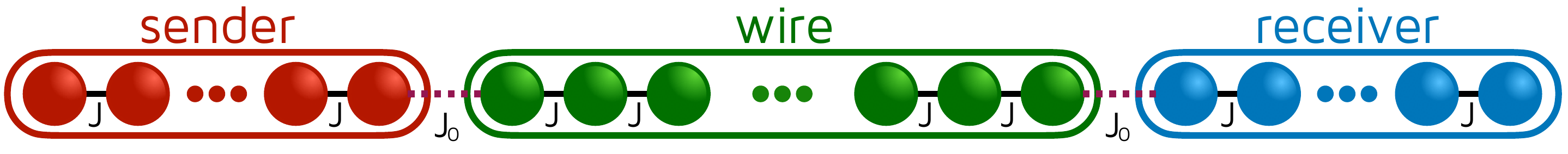}
	\caption{Setup of an $n$ interacting qubit quantum state transfer protocol ($n$-iQST). The sender and receiver blocks are weakly coupled by $J_0$ at both edges of a wire. Each part consists of a 1D lattice described by the Hamiltonian in Eq.~\eqref{Eq:Ham1} with i) nearest neighbor couplings only, ii) $\Delta=0$, iii) uniform parameters (that is, $J_i=J$ and $h_i=h$ $\forall i$) and iv) $J_0\ll J$.}
	\label{F.Figure}
\end{figure}

Here, we propose a high-quality 3- and 4-iQST protocol by means of an integrable $U(1)$-symmetric Hamiltonian (see Eq.~\eqref{E_Ham_Hei}) based on a weak-coupling protocol between the sender (receiver) blocks and the wire.
We consider a 1D spin-$\frac{1}{2}$ chain with isotropic interactions in the $XY$ plane
\begin{align}\label{Eq:Ham1}
\hat{H} = \frac{1}{4}\sum_{i}^{N}J_i\left(\hat{\sigma}^{x}_{i}\hat{\sigma}^{x}_{i+1} + \hat{\sigma}^{y}_{i}\hat{\sigma}^{y}_{i+1}\right)+\frac{h_i}{2}\hat{\sigma}_i^z~,
\end{align}
where $\hat{\sigma}^{\alpha}_{i}$ ($\alpha=x,y,z$) is the Pauli operator sitting on site $i$, and we assume open boundary conditions $\hat{\sigma}^{\alpha}_{N+1}=0$.
In the following, we will also assume that the couplings $J_i$ are all uniform, except for the couplings $J_i=J_0$ between the sender (receiver) block and the wire (see Fig.~\ref{F.Figure}). This is the so-called weak-coupling scheme which has been already successfully investigated for 1- and 2-qubit QST~\cite{Wojcik2005,Lorenzo2015,Lorenzo2017b,Apollaro2020a}. We will also set the coupling within the sender (receiver) block and within the wire as our time and energy unit $J_i=J=1$. Note that these assumptions are unnecessary for the diagonalisation of the model we are going to outline.

%
%

Using the Jordan-Wigner transformation, Eq.~\eqref{Eq:Ham1} is mapped to a spinless quadratic fermion model~\cite{LIEB1961407},
\begin{align}\label{Eq:Ham4}
\hat{H} = \sum_{i}^{N}\frac{J_i}{2}\left(\hat{c}^{\dagger}_{i}\hat{c}_{i+1} + \text{h.c.}\right)+\sum_{i}^{N} h_i \hat{c}^{\dagger}_i\hat{c}_i-\sum_{i}^{N}\frac{h_i}{2}~,
\end{align}
where, hereafter, the energy is rescaled by the constant term.

The $U(1)$ symmetry of the model implies that the number operator, $\hat{\mathcal{N}}=\sum_{i=1}^N \hat{c}_i^{\dagger}\hat{c}_i$ is a conserved quantity. This allows the dynamics to be addressed in excitation-number invariant subspaces. Moreover, due to the quadratic, i.e., non-interacting, nature of the Hamiltonian, the $n$-qubit dynamics can be expressed in terms of the single-particle transition amplitudes.
In the single-particle sector, Eq.~\eqref{Eq:Ham4} is diagonalised as
\begin{align}\label{E_1part}
\hat{H}=\sum_{k=1}^N \omega_k \ket{\phi_k}\!\!\bra{\phi_k}\equiv \sum_{k=1}^N \omega_k \hat{c}_k^{\dagger}\hat{c}_k~,
\end{align}
where $\{\omega_k,\ket{\phi_k}\}$, with $ \ket{\phi_k}=\hat{c}_k^{\dagger}\ket{0}$, are the eigenvalues and the eigenvectors of the tridiagonal matrix, $A\equiv \bra{i}\hat{H}\ket{j}=\frac{J_i}{2}\left(\delta_{i,j+1}+\delta_{i,j-1}\right)+h_i \delta_{i,j}$, describing the single-particle dynamics in the direct space basis, $\ket{i}\equiv \hat{c}_i^{\dagger}\ket{\mathbf{0}}$~.

Finally, the single-particle transition amplitude from site $i$ to to site $j$ reads
\begin{align}
\label{E.Sparticel_amp}
f_i^j(t)=\bra{j}e^{-it \hat{H}}\ket{i}=\sum_{k=1}^{N}e^{-i \omega_k t }\bra{j}\phi_k\rangle\langle \phi_k\ket{i}=\sum_{k=1}^{N}e^{-i \omega_k t }\phi_{jk}\phi_{ki}~,
\end{align}
and builds up the transition amplitude hermitian matrix
\begin{align}\label{E_FMatrix}
\mathcal{F}(t)=
\begin{pmatrix}
f_1^1(t) & f_1^2(t) & \cdots & f_1^N(t)\\
f_2^1(t) & f_2^2(t) & \cdots & f_2^N(t)\\
\vdots & \vdots & \ddots &  \vdots\\
f_N^1(t) & f_N^2(t) & \cdots & f_N^N(t)\\
\end{pmatrix}~.
\end{align}

The transition amplitude for the transfer of $n_s$ excitations, residing on the sender sites $\{n_s\}=\{s_1,s_2,\dots,s_{n_s}\}$, to the receiver sites $r$, residing on the receiver sites $\{n_r\}=\{r_1,r_2,\dots,r_{n_r}\}$, is the minor $\mathcal{F}(t)_{\{n_s\}}^{\{n_r\}}$ of $\mathcal{F}(t)$, i.e., the determinant of the matrix where only the
$\{n_s\}$ rows and  $\{n_r\}$ columns of $\mathcal{F}(t)$ are considered. The minor $\{n_r\}=\{r_1,r_2,\dots,r_{n_r}\}$ is the quantity of interest entering Eq.~\eqref{E_Fn_1} where it is represented by $f_S^S$.

\subsection{3-iQST via weak links}\label{sS.3qst_spin}
From the previous discussion, we derive that, in order to have $\average{F_3}=1$, each of the transition amplitudes $f_i^j$ belonging to distinct sets of Eq.~\eqref{E_Fn_1}, i.e., those belonging to a different row and column of Eq.~\eqref{E_FMatrix}, need to have unit modulus and the same phase.

\begin{figure}[!ht]
    \centering
    \includegraphics[width =1\textwidth]{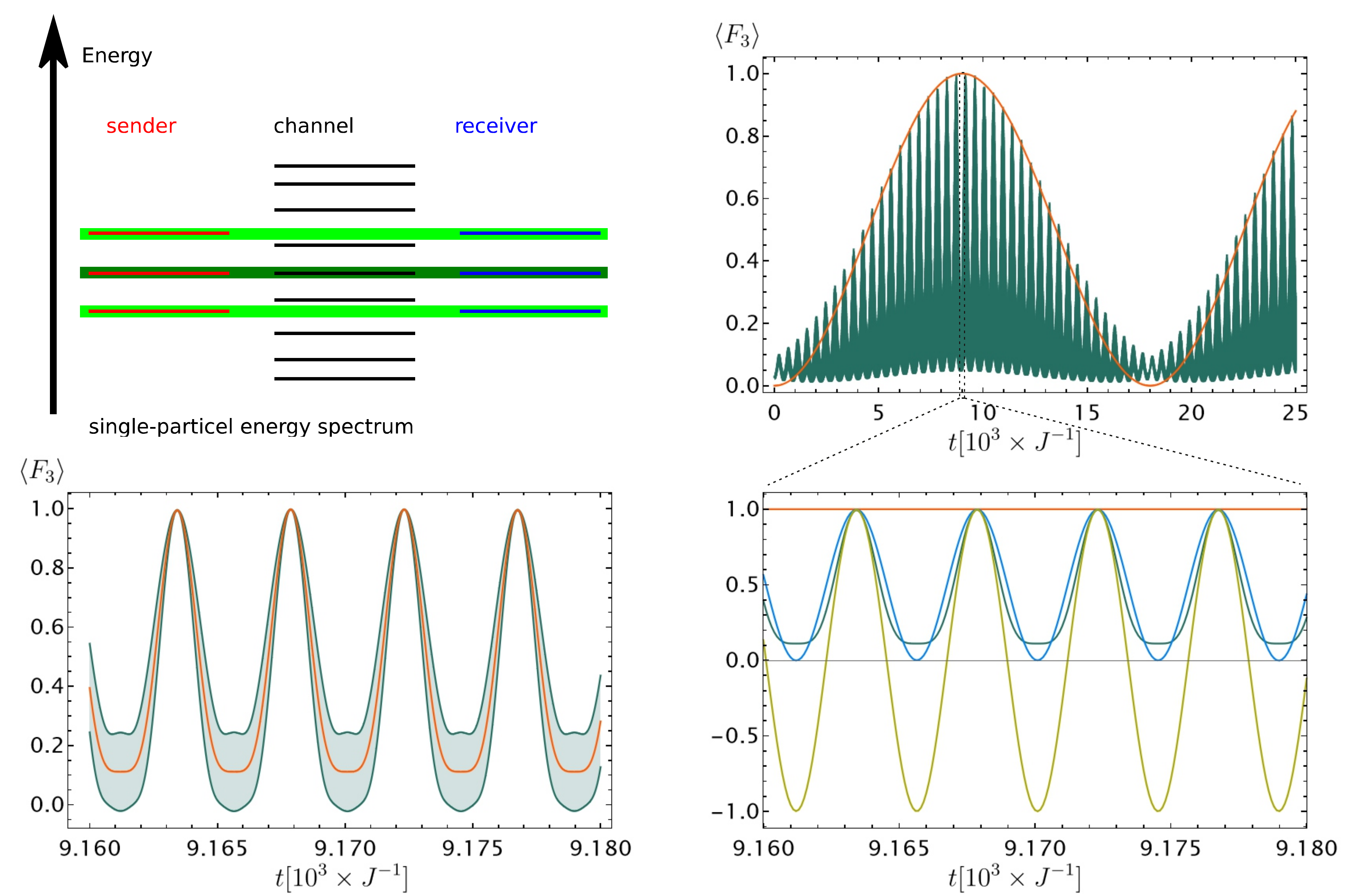}
	\caption{(upper left) Resonance conditions for 3-qubits iQST via a $n_w=9$ site wire for a total length of $N=15$ chain with $J_0=0.01$. The central, dark-green band indicates the \nth{1}-order perturbative correction to the energy levels and the external, light-green bands the \nth{2}-order ones. (upper right) 3-iQST average fidelity $\average{F_3}$ vs $t$ (green line) and the $\average{F}_{\text{env}}$ given by the sinusoidal function (orange line); (lower right) shows a zoom for the average fidelity $\average{F}_3$ (green), the single-particle transition amplitudes $f_1^1=f_3^3$ (blue) and $f_2^2$ (goldenrod). (lower left) Average fidelity with its standard deviation $\average{F_3}\pm\sigma$. }
	\label{fig:fid3}
\end{figure}

Without loss of generality, we choose $i-j$ (or, equivalently, $i+j$) to be even, so that $f_i^j$ is purely real (imaginary) by choosing the transition amplitudes between sites $\left(1,N-2\right),\left(2,N-1\right),\left(3,N\right)$ for $N$ odd (even). From Ref.~\cite{Chetcuti2020}, we know that for a length of the wire $n_w=4l+1$ and $n_w=4l+3$ there are, respectively, one and three resonant single-particle levels with the sender (receiver) block. We will refer to these two cases as non-resonant and resonant 3-iQST respectively, because, in the former case, the QST time is ruled by the non-resonant energy level splitting and in the latter by the resonant energy level splitting.
For the non-resonant case, i.e., for length of the wire $n_w=4l+1$, each single-particle transition amplitude in Eq.~\eqref{E.Sparticel_amp} can be approximated, up to \nth{2}-order, by
\begin{align}
	\label{E_fnr}
	f_{i}^{j}(t)=\sum_{k=1}^{7}e^{-i \omega_k t }\phi_{jk}\phi_{ki}~,
\end{align}
where only the quasi-degenerate energy levels enter the sum, 
and occupy, in the increasing ordered energy spectrum $\omega_k<\omega_{k'}$ for $k<k'$, the following positions: the four \nth{2}-order perturbed energy levels are at $\{\lceil \frac{N-5}{4}\rceil ,\lceil \frac{N-5}{4}\rceil+1,\lceil \frac{N-5}{4}\rceil+n_w,\lceil \frac{N-5}{4}\rceil+n_w+1\}$, where $\lceil x\rceil$ is the ceiling function, and the three \nth{1}-order perturbed energy levels are at $\left\{\frac{N+1}{2}-1,\frac{N+1}{2},\frac{N+1}{2}+1\right\}$. In Fig.~\ref{fig:fid3}, an instance of the single-particle energy levels is given for the non-resonant case with $n_w=9$. Exploiting the parity relations for the eigenvectors of mirror-symmetric matrices~\cite{Banchi2013}, $\phi_{k,N+1-i}=\left(-1\right)^k \phi_{k,N+1-i}$, and elementary trigonometric identities, it is easy to show that the longest time-scale is governed by the \nth{2}-order perturbative energy splitting. As a consequence, the envelope of the $n$-qubit QST average fidelity is given by $\average{F}_{\text{env}}\simeq \left|\sin^2 \left(\frac{\delta \omega }{2}t\right)\right|^2$, where $\delta \omega=\omega_{\lceil \frac{N-5}{4}\rceil+1}-\omega_{\lceil \frac{N-5}{4}\rceil}$. Within the transfer time $\tau=\frac{\pi}{\delta \omega}$, oscillations on a timescale of order of $J$ occur because of the internal dynamics of the receiver block. Nevertheless, the fidelity reaches its maximum value of $\average{F}=1-O\left(J_0^2\right)$ multiple times, giving the receiver the opportunity to read-out the state within a time-window of the order of $J$. In Fig.~\ref{fig:fid3} we show an instance of the aforementioned timescales.

\subsection{4-iQST via weak links}\label{sS.4qst_spin}
\begin{figure}[!ht]
    \centering
	\includegraphics[width=0.9\textwidth]{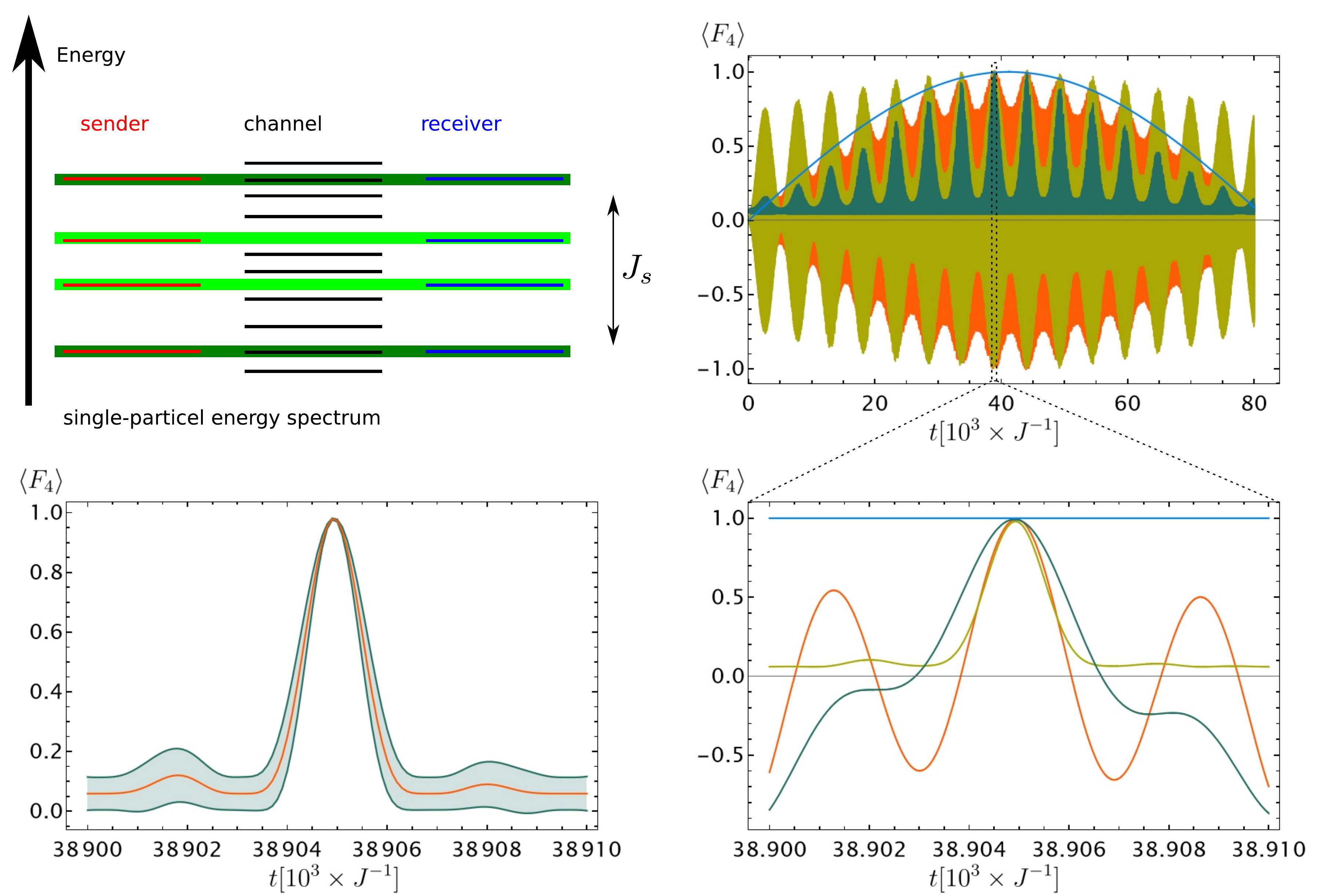}
	\caption{$4$-QST average fidelity vs. time $t$ via a quantum channel with $n_w=10$ spins. (upper right)  Resonance conditions for 4-qubits iQST via a $n_w=10$ site wire for a total length of $N=18$ chain with $J_0=0.01$. The two central, light-green bands indicates the \nth{1}-order perturbative correction to the energy levels and the external, dark-green bands the \nth{2}-order ones. This resonance condition can be achieved by acting on the sender (receiver) coulings $J_S$ as given in Eq.~\eqref{E_4res}. (upper-right) Average fidelity $\average{F_4}$ (dark blue), its sinusoidal envelop (blue), and the single-particle transition amplitudes $f_1^1=f_4^4$ (orange) and $f_2^2=f_3^3$ (goldenrod). (lower right) Zoom of the dotted box in the upper right panel with the same color code. (lower left) Average fidelity with its standard deviation $\average{F_4}\pm\sigma$. }
	\label{fig:fid4}
\end{figure}

Here we turn our attention to the QST of 4 interacting qubits, again over the channel depicted in Fig.~\ref{F.Figure}, and with the same Hamiltonian as in Eq.~\eqref{Eq:Ham1}, where, now, the weak-coupling condition entails $J_i=J_0\left(\delta_{i,4}+\delta_{i,N-4}\right)$ and we set $J_i=1$ otherwise. However, it has been shown in Ref.~\cite{Chetcuti2020} that, for such a uniform coupling scheme, either all, or none, of the four single-particle energy levels are in resonance with the wire's energy level. As a consequence, because of the incommensurability of the frequencies entering Eq.~\eqref{E_fnr} (where now the summations extends over 12 or 8 frequencies for the all-resonant and the non-resonant cases, respectively), the average fidelity of the 4-iQST will not approach unity. In order to introduce a time-scale separation, as done in the 3-iQST case, a minimal-engineering solution can be achieved by acting on the intra-sender (-receiver) couplings $J_i$  such that two (symmetric) energy levels are in resonance with the wire's energies and two are left out-of-resonance. The values for $J_s$ can be readily found by setting
\begin{align}
	\label{E_4res}
	J_s=\frac{\cos\frac{k\pi}{n_w+1}}{\cos\frac{s\pi}{5}}~,
\end{align}
where $s\in\left(1,2\right)$ represents the $s^{th}$-energy level of the sender put in resonance with the $k^{th}$-energy level of the wire for that value of $J_s$. In Fig.~\ref{fig:fid4} a schematic representation is given for $n_w=10$, $k=2$, and $s=1$. With such a protocol, we achieve a separation of time-scales in the single-particle transition amplitudes, and achieve, once again, a transfer time of the order of magnitude of the \nth{2}-order energy perturbation correction, represented in Fig.~\ref{fig:fid4} by the sinusoidal envelop of $\bar{F_4}$. Without loss of generality, we report in Fig.~\ref{fig:fid4} the single-particle transition amplitudes $f_1^1=f_4^4\simeq f_2^2=f_3^3>0.99$, resulting in $\average{F_4}\simeq 0.98$ at the optimal transfer time.

\subsection{Comparison between \textit{n}-QST and \textit{n}-iQST}
In Sec.~\ref{sectre} and Sec.~\ref{S.XX}, we showed that a single interacting channel is able to efficiently transfer the quantum state of $n$ spins, with $n$ independent (parallel) transmissions if they are non-interacting, and in a single transfer instance, if we are in the presence of interactions between the sender's qubits, up to $n=4$.
Here, we address the question: which of these two type of transfer performs more reliably at a fixed value of given average fidelity? We do this by using the variance in Eq.~\eqref{E_f2} as a figure of merit. 

\begin{figure}[!ht]
    \centering
	\includegraphics[width =0.49\textwidth]{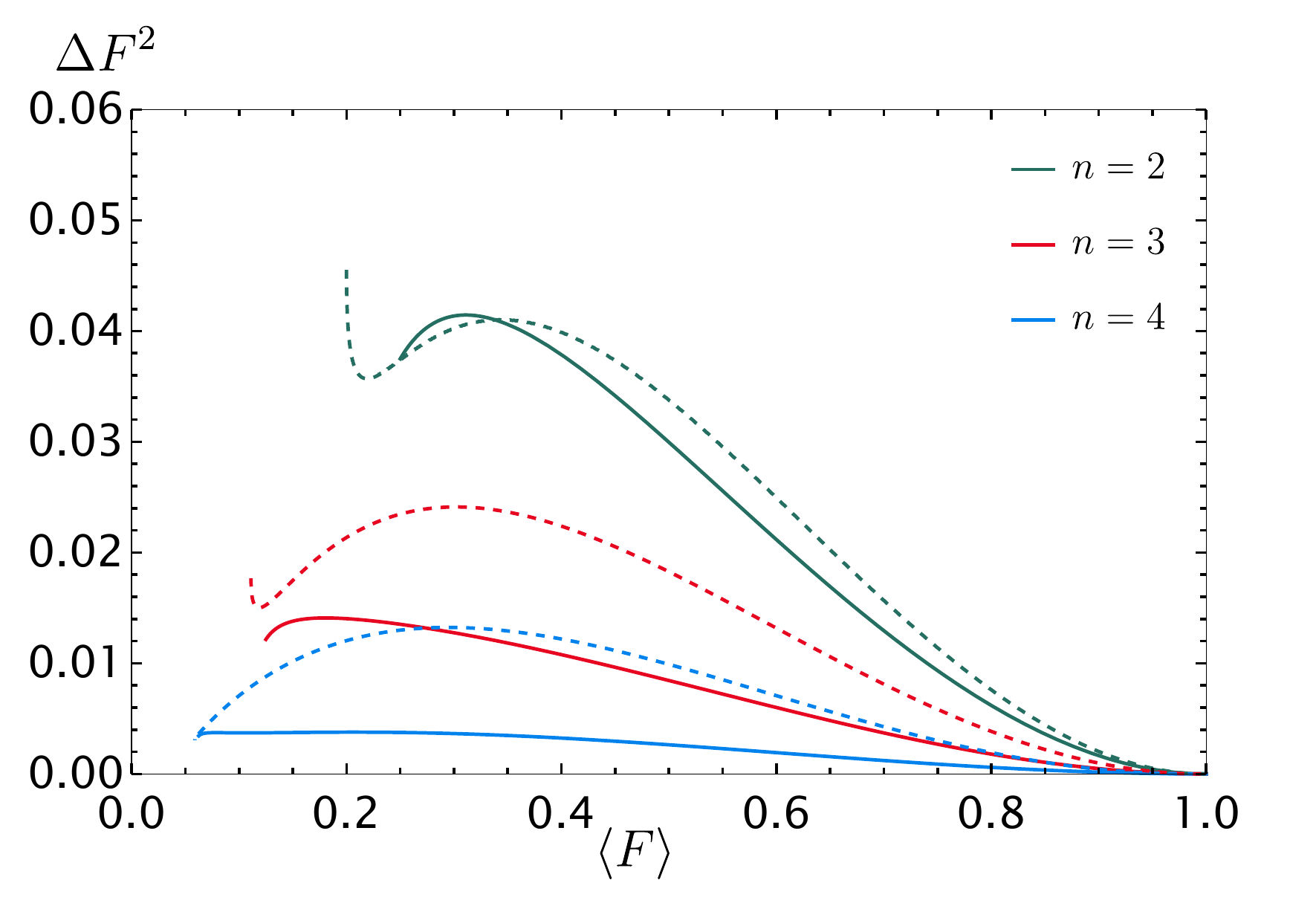}
	\caption{Variance $\Delta F^2$ vs average Fidelity $\average{F}$ for independent channels (continuous lines), as depicted in Fig.~\ref{F.Figure_ind}, and for an interacting channel (dotted lines), as depicted in Fig.~\ref{F.Figure}. Curves are for $n=2,3,4$, respectively green, red, and blue.}
	\label{fig_9}
\end{figure}

As can be seen from Fig.~\ref{fig_9}, at fixed average fidelity $\average{F}$, the variance $\Delta F^2$ is always greater for the case of $n$-iQST with a single channel, than for $n$-QST with $n$ independent channels. This can be readily explained by noticing that the presence of interactions allows the receivers to explore a higher portion of the Hilbert space during the evolution, with respect to the case of independent channels. E.g., in the case of independent channels, the entanglement of the receivers can not exceed that of the senders, whereas this is not the case for the single channel scenario.

\section{Concluding remarks}\label{secconclu}
Quantum state transfer of an $n$-qubit system is a key protocol in many quantum information processing tasks. Whereas single qubit quantum state transfer has been intensively investigated, and also experimentally realised on a variety of different experimental platforms, arbitrary $n$-qubit quantum state transfer is still a goal to be achieved. The quantum transfer of a many-body interacting system is a formidable task, made difficult both by the exponentially-increasing dimensionality of the Hilbert space, and by the complexity due to the particle interactions.
In this paper, we have provided a new approach to the $n$ interacting qubit QST via dynamical maps. By considering the receiver block as an open quantum system coupled to an environment, embodied both by the sender block and the quantum channel, we have derived a general expression for the average fidelity of $n$-iQST in terms of quantum dynamical maps elements. We have also analysed the dispersion of the values of the fidelity, when evaluated on all possible pure input states, that we expressed by the fidelity variance, for which we provided a general expression as well. Then, we specialised to short-range transmission obtained by coupling senders and receivers to a linear spin chain. In the case of $U(1)$-symmetric dynamics, we expressed the average fidelity in terms of transition amplitudes in the occupation-number invariant subspaces. We investigated in detail the $U(1)$-symmetric $XX$ spin-$\frac{1}{2}$ Hamiltonian, and, by exploiting its non-interacting nature in the fermionic representation, we were able to express the $n$-qubit iQST average fidelity only in terms of single-particle transition amplitudes. 

Our formalism also encompasses the non-interacting scenario, where we have shown that independent channels achieve a higher $n$-QST fidelity, at a fixed single-particle transition amplitude, for product states than for entangled states, although the variance of the former is greater than that of the latter, at a fixed value of the average fidelity. Interestingly, we obtained that the average fidelity is a self-averaging quantity, as quantified by the vanishing coefficient of variation for $n\gg 1$ at high-values of the average fidelity.

Finally, we have proposed a protocol for the high-fidelity transfer of the quantum state of both 3 and 4 interacting qubits, arranged in a linear chain with uniform couplings, via a weak-coupling scheme to a non-engineered $XX$ spin-$\frac{1}{2}$ chain. Whereas up to $n=4$, high-quality iQST can be achieved by means of a uniform channel, it appears that for $n>4$ our protocol has to be substantially modified because of the impossibility to introduce a time-scale separation in the single-particle transition amplitudes 
when the involved energy levels become too numerous. 

Considering the importance of quantum state transfer of a many-qubit system in several quantum information processing tasks, ranging from cryptography to quantum computation, it is crucial to establish a theoretical framework that can encompass all of the various possibilities. We believe that this can be provided by the quantum map approach supplemented by the investigation of the statistics of the fidelity, that we started to establish in this paper, and that we applied to the general $n$-qubit state transfer. Moreover, our approach explicitly includes interactions among the sender's particles, opening the way to investigate quantum transfer protocols of complex interacting systems.

\section*{Acknowledgemts}
TJGA thanks Abolfazl Bayat for pointing him out to Ref.~\cite{Bayat2007}, where useful expressions for the $d$-level fidelity can be found. SL acknowledges support from MIUR through project PRIN Project 2017SRN-BRK QUSHIP. TJGA ackowledges funding through the IPAS+ (Internationalisation Partnership Awards Scheme +) ACROSS project by the MCST (The Malta Council for Science \& Technology).
We also thanks the organizers of the Quantum Hiking Conference in 2019 where useful discussions concerning this work have taken place.

\appendix

\section{1-qubit map}\label{appendix1}
The 1-qubit density matrix map $\hat{\rho}_N=\Phi(t)\hat{\rho}_1$ is given by
\begin{align}\label{E_1ex_map}
\begin{pmatrix}
\rho_{00}\\
\rho_{01}\\
\rho_{10}\\
\rho_{11}\\
\end{pmatrix}_N=
\begin{pmatrix}
1 &0&0&1-\left|f_1^N\right|^2\\
0&f_1^N&0&0\\
0&0&\left(f_1^N\right)^*&0\\
0&0&0&\left|f_1^N\right|^2
\end{pmatrix}
\begin{pmatrix}
\rho_{00}\\
\rho_{01}\\
\rho_{10}\\
\rho_{11}\\
\end{pmatrix}_1
\end{align}

\section{2-qubit map}\label{appendix2}

From Ref.~\cite{Apollaro2015}, we derive the following two-qubit map's elements $\hat{\rho}_{N-1,N}=\Phi(t)\hat{\rho}_{1,2}$,
\begin{align}
\label{E_2map}
&A_{00}^{00}=1~,~A_{00}^{11}=1-\left|f_1^{N-1}\right|^2-\left|f_1^{N}\right|^2~,~A_{00}^{22}=1-\left|f_2^{N-1}\right|^2-\left|f_2^{N}\right|^2~,~\nonumber\\
&A_{00}^{33}=1-\left|f_{12}^{mN-1}\right|^2-\left|f_{12}^{mN}\right|^2-\left|f_{12}^{N-1N}\right|^2~,~ \nonumber\\ &A_{00}^{12}=-f_1^{N-1}\left(f_2^{N-1}\right)^*-f_1^{N}\left(f_2^{N}\right)^*~,~
A_{00}^{21}=-f_2^{N-1}\left(f_1^{N-1}\right)^*-f_2^{N}\left(f_1^{N}\right)^*~,~\nonumber\\
&A_{01}^{01}=\left(f_1^N\right)^*~,~A_{01}^{02}=\left(f_2^N\right)^*~,~A_{01}^{13}=f_1^m\left(f_{12}^{mN}\right)^*~,~A_{01}^{23}=f_2^m\left(f_{12}^{mN}\right)^*~,~\nonumber\\
&A_{02}^{01}=\left(f_{1}^{N-1}\right)^*~,~A_{02}^{02}=\left(f_{2}^{N-1}\right)^*~,~A_{02}^{13}=f_1^m\left(f_{12}^{mN-1}\right)^*~,~A_{02}^{23}=f_2^m\left(f_{12}^{mN-1}\right)^*~,~\nonumber\\
&A_{03}^{03}=\left(f_{12}^{N-1N}\right)^*~,~\nonumber\\
&A_{11}^{11}=\left|f_1^N\right|^2~,~A_{11}^{12}=f_1^N\left(f_2^N\right)^*~,~A_{11}^{21}=f_2^N\left(f_1^N\right)^*~,~A_{11}^{22}=\left|f_2^N\right|^2~,~A_{11}^{33}=\left|f_{12}^{mN}\right|^2~,~\nonumber\\
&A_{12}^{11}=f_1^N\left(f_1^{N-1}\right)^*~,~A_{12}^{12}=f_1^N\left(f_2^{N-1}\right)^*~,~
A_{12}^{21}=f_2^N\left(f_1^{N-1}\right)^*~,~A_{12}^{22}=f_2^N\left(f_2^{N-1}\right)^*~,~\nonumber\\
&A_{12}^{33}=f_{12}^{mN}\left(f_{12}^{mN-1}\right)^*~,~\nonumber\\
&A_{13}^{13}=f_1^N\left(f_{12}^{N-1N}\right)^*~,~A_{13}^{23}=f_2^N\left(f_{12}^{N-1N}\right)^*~,~\nonumber\\
&A_{22}^{33}=\left|f_{12}^{mN-1}\right|^2~,~A_{22}^{11}=\left|f_{1}^{N-1}\right|^2~,~A_{22}^{22}=\left|f_{2}^{N-1}\right|^2~,~A_{22}^{12}=f_{1}^{N-1}\left(f_{2}^{N-1}\right)^*~,~\nonumber\\
&A_{22}^{21}=f_{2}^{N-1}\left(f_{1}^{N-1}\right)^*~,~\nonumber\\
&A_{23}^{13}=f_{1}^{N-1}\left(f_{12}^{N-1N}\right)^*~,~A_{23}^{23}=f_{2}^{N-1}\left(f_{12}^{N-1N}\right)^*~,~A_{33}^{33}=\left|f_{12}^{N-1N}\right|^2~,
\end{align}
where $m$ denotes the summation over all $i\neq S,R$.

\begin{align*}\label{E_2ex_map}
\scriptsize
\begin{pmatrix}
\rho_{00}\\
\rho_{01}\\
\rho_{02}\\
\rho_{03}\\
\rho_{10}\\
\rho_{11}\\
\rho_{12}\\
\rho_{13}\\
\rho_{20}\\
\rho_{21}\\
\rho_{22}\\
\rho_{23}\\
\rho_{30}\\
\rho_{31}\\
\rho_{32}\\
\rho_{33}\\
\end{pmatrix}_{N-1,N} \hspace{-1.5em} =
\begin{pmatrix}
A_{00}^{00}& 0 & 0 & 0 & 0 & A_{00}^{11} & A_{00}^{12} & 0 & 0 & A_{00}^{21} & A_{00}^{22} & 0 & 0 & 0 & 0 & A_{00}^{33} \\
0& A_{01}^{01} & A_{01}^{02} & 0 & 0 & 0 & 0 & A_{01}^{13} & 0 & 0 & 0 & A_{01}^{23} & 0 & 0 & 0 & 0 \\
0 & A_{02}^{01} & A_{02}^{02} & 0 & 0 & 0 & 0 & A_{02}^{13} & 0 & 0 & 0 & A_{02}^{23} & 0 & 0 & 0 & 0 \\
0& 0 & 0 & A_{03}^{03} & 0 & 0 & 0 & 0 & 0 & 0 & 0 & 0 & 0 & 0 & 0 & 0 \\
0& 0 & 0 & 0 & A_{10}^{10} &  0  & 0 & 0 & A_{10}^{20} & 0 & 0  & 0 &0  & A_{10}^{31} & A_{10}^{32} & 0 \\
0& 0 & 0 & 0 & 0 & A_{11}^{11} & A_{11}^{12} & 0 & 0 & A_{11}^{21} & A_{11}^{22} & 0 & 0 & 0 & 0 & A_{11}^{33} \\
0& 0 & 0 & 0 & 0 & A_{12}^{11} & A_{12}^{12} & 0 & 0 & A_{12}^{21} & A_{12}^{22} & 0 & 0 & 0 & 0 & A_{12}^{33} \\
0& 0 & 0 & 0 & 0 & 0 & 0 & A_{13}^{13} & 0 & 0 & 0 & A_{13}^{23} & 0 & 0 & 0 & 0 \\
0& 0 & 0 & 0 &  A_{20}^{10} & 0 & 0 &  0 & A_{20}^{20} & 0 & 0 & 0 & 0 & A_{20}^{31} & A_{20}^{32} & 0 \\
0& 0 & 0 & 0 & 0 & A_{21}^{11} & A_{21}^{12} & 0 & 0 & A_{21}^{21} & A_{21}^{22} & 0 & 0 & 0 & 0 &  A_{21}^{33} \\
0& 0 & 0 & 0 & 0 & A_{22}^{11} & A_{22}^{12} & 0 & 0 & A_{22}^{21} & A_{22}^{22} & 0 & 0 & 0 & 0 & A_{22}^{33} \\
0& 0 & 0 & 0 & 0 & 0 & 0 & A_{23}^{13} & 0 & 0 & 0 & A_{23}^{23} & 0 & 0 & 0 & 0\\
0& 0 & 0 & 0 & 0 & 0 & 0 & 0 & 0 & 0 & 0 & 0 & A_{30}^{30} & 0 & 0 &  0\\
0& 0 & 0 & 0 & 0 & 0 & 0 & 0 & 0 & 0 & 0 & 0 & 0 & A_{31}^{31} & A_{31}^{32} &0 \\
0& 0 & 0 & 0 & 0 & 0 & 0 & 0 & 0 & 0 & 0 & 0 & 0 & A_{32}^{31} & A_{32}^{32} & 0 \\
0& 0 & 0 & 0 & 0 & 0 & 0 & 0 & 0 & 0 & 0 & 0 & 0 & 0 & 0 & A_{33}^{33}
\end{pmatrix}
\begin{pmatrix}
\rho_{00}\\
\rho_{01}\\
\rho_{02}\\
\rho_{03}\\
\rho_{10}\\
\rho_{11}\\
\rho_{12}\\
\rho_{13}\\
\rho_{20}\\
\rho_{21}\\
\rho_{22}\\
\rho_{23}\\
\rho_{30}\\
\rho_{31}\\
\rho_{32}\\
\rho_{33}\\
\end{pmatrix}_{1,2}
\end{align*}

\section*{References}

\bibliographystyle{unsrt.bst}
\bibliography{nqst.bib}

\end{document}